\documentclass[journal=jctcce,manuscript=article]{achemso}
\setkeys{acs}{articletitle = true}
\usepackage[version=3]{mhchem}
\usepackage[utf8]{inputenc}
\usepackage{color}
\usepackage{url}
\usepackage[shortlabels]{enumitem}
\usepackage{bm}
\usepackage{subfig}
\usepackage{xfrac}
\usepackage{mathtools, cuted}
\usepackage{atbegshi}
\AtBeginDocument{\AtBeginShipoutNext{\AtBeginShipoutDiscard}}

\newcommand{\br}{\mathbf{r}}
\newcommand{\bR}{\mathbf{R}}

\newcommand{\bra}[1]{\langle#1|}
\newcommand{\ket}[1]{|#1\rangle}
\newcommand{\Tr}{\mathrm{Tr}}
\newcommand{\im}{\mathrm{i}}
\newcommand{\invs}{S^{-1}}
\newcommand{\bF}{\mathbf{F}}
\newcommand{\bE}{\mathbf{E}}

\newcommand{\bmu}{\bm{\mu}}
\newcommand{\bpol}{\bm{\alpha}}
\newcommand{\be}{\mathbf{\hat{e}}}
\newcommand{\Ef}{\mathcal{E}}

\newcommand*{\revone}[1]{{\color{black} #1}}
\newcommand*{\revtwo}[1]{{\color{black} #1}}

\author{Franco P. Bonaf\'e}
\affiliation{Max Planck Institute for the Structure and Dynamics of Matter. Hamburg, Germany}
\alsoaffiliation{Universidad Nacional de C\'ordoba. Facultad de Ciencias Qu\'imicas, Departamento de Qu\'imica Te\'orica y Computacional. C\'ordoba, Argentina}
\alsoaffiliation{Instituto de Investigaciones en Fisicoqu\'imica de C\'ordoba, INFIQC (CONICET - Universidad Nacional de C\'ordoba). C\'ordoba, Argentina}
\author{B\'alint Aradi}
\affiliation{Bremen Center for Computational Materials Science, Universität Bremen. Bremen, Germany}
\author{Ben Hourahine}
\affiliation{SUPA, Department of Physics, John Anderson Building, The University of Strathclyde, 107 Rottenrow, Glasgow G15 6QN, United Kingdom}
\author{Carlos R. Medrano}
\affiliation{Universidad Nacional de C\'ordoba. Facultad de Ciencias Qu\'imicas, Departamento de Qu\'imica Te\'orica y Computacional. C\'ordoba, Argentina}
\alsoaffiliation{Instituto de Investigaciones en Fisicoqu\'imica de C\'ordoba, INFIQC (CONICET - Universidad Nacional de C\'ordoba). C\'ordoba, Argentina}
\author{Federico J. Hern\'andez}
\affiliation{Department of Physics, Universidad de Santiago de Chile, Av. Ecuador 3493, Santiago, Chile}
\alsoaffiliation{Universidad Nacional de C\'ordoba. Facultad de Ciencias Qu\'imicas, Departamento de Qu\'imica Te\'orica y Computacional. C\'ordoba, Argentina}
\alsoaffiliation{Instituto de Investigaciones en Fisicoqu\'imica de C\'ordoba, INFIQC (CONICET - Universidad Nacional de C\'ordoba). C\'ordoba, Argentina}
\author{Thomas Frauenheim}
\affiliation{Bremen Center for Computational Materials Science, Universität Bremen. Bremen, Germany}
\alsoaffiliation{Computational Science Research Center (CSRC) Beijing and Computational Science and Applied Research (CSAR) Institute Shenzhen, China}
\author{Cristi\'an G. S\'anchez}
\affiliation{Instituto Interdisciplinario de Ciencias Básicas, Universidad Nacional de Cuyo, CONICET, Facultad de Ciencias Exactas y Naturales, Mendoza, Argentina}
\email{csanchez@mendoza-conicet.gob.ar}

\title{A real-time time-dependent density functional tight-binding implementation for semiclassical excited state electron-nuclear dynamics and pump-probe spectroscopy simulations}

\begin{document}

\begin{tocentry}
\centering
\includegraphics[width=0.65\linewidth]{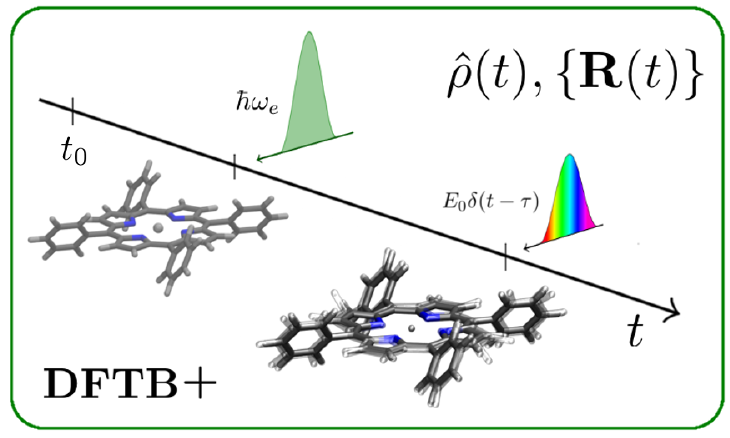}
\end{tocentry}

\newpage
\begin{abstract}
The increasing need to simulate the dynamics of photoexcited molecular and nanosystems in the sub-picosecond regime demands new efficient tools able to describe the quantum nature of matter at a low computational cost. By combining the power of the approximate DFTB method with the semiclassical Ehrenfest method for nuclear-electron dynamics we have achieved a real-time time-dependent DFTB (TD-DFTB) implementation that fits such requirements. In addition to enabling the study of nuclear motion effects in photoinduced charge transfer processes, our code adds novel features to the realm of static and time-resolved computational spectroscopies. In particular, the optical properties of periodic materials such as graphene nanoribbons or the use of corrections such as the ``LDA+U'' and ``pseudo SIC'' methods to improve the optical properties in some systems, can now be handled at the TD-DFTB level. Moreover, the simulation of fully-atomistic time-resolved transient absorption spectra and impulsive vibrational spectra can now be achieved within reasonable computing time, owing to the good performance of the implementation and a parallel simulation protocol. Its application to the study of UV/visible light-induced vibrational coherences in molecules is demonstrated and opens a new door into the mechanisms of non-equilibrium ultrafast phenomena in countless materials with relevant applications.
\end{abstract}

\section{Introduction}

Harnessing the excited state properties of molecules and nanostructures has become one of the main paths to successful solutions for technological challenges, ranging from artificial photosynthesis and solar cells to photocatalysis. It is also the core idea behind many fundamental lines of research in photochemistry, pump-probe spectroscopy, scanning tunneling microscopy, and so on. The large amount of research in these fields pushes the development of novel computational tools to efficiently perform simulations of these relevant processes~\cite{Curchod2013,Crespo-Otero2018}. The most general solution would consist of solving the time-dependent Schr\"odinger equation (TDSE) for coupled electronic and nuclei wavefunctions, thus considering both subsystems as quantum entities. However this approach turns out to be too computationally demanding for many applications and approximations have to be made.

Mixed quantum-classical approaches, which consider the nuclei as classical while keeping the electronic sub-system quantum, provide a way to achieve feasible simulations of excited state processes. The most commonly used trajectory-based methods are surface hopping and Ehrenfest dynamics~\cite{Hack2000,Curchod2013,Crespo-Otero2018}. In the former, the potential energy surfaces (PES) of the system are calculated ``on the fly'' and in each trajectory stochastic transitions between PES are allowed based on the flux of population between pairs of states, giving rise to the fewest-switches surface hopping (FSSH) method, one of its many variants. In FSSH, an ensemble of semi-classical trajectories must be run in order to compute the dynamics of interest~\cite{Tully1990}. In Ehrenfest dynamics, on the other hand, the system is allowed to evolve in an average PES that results from a combination of all the occupied adiabatic states, in contrast to FSSH in which dynamics on a single PES is well described in uncoupled regions~\cite{Curchod2013}.

FSSH has been widely used to study transition probabilities along different paths to explain phenomena such as photo-chemical reactions~\cite{Tapavicza2007}, where the Ehrenfest method normally fails due to not allowing the system to hop between different PES. However, while Ehrenfest dynamics can be derived from first principles, by introducing the ansatz of a single-product of the total wavefunction in the TDSE~\cite{Curchod2013}, the FSSH can not be derived from first principles, although several quasi-derivations have been attempted~\cite{Subotnik2013,Kapral2016}.

While several implementations of Ehrenfest dynamics have been published for model or atomistic systems, within different formalisms, there is a clear need for publicly available programs that propagate electron-nuclear trajectories for a wide range of atomistic systems using accurate yet efficient formalisms. Implementations of open source, free-software offering real-time propagation of time-dependent density functional theory (TD-DFT) scheme include the \textsc{octopus} code~\cite{Andrade2015a,Tancogne-Dejean2020}. Other implementations of Ehrenfest TD-DFT have also been reported~\cite{Wang2011h,Meng2008b}.For some applications, however, it is useful to apply a more approximate formalism that, in turn, allows one to tackle very large systems or reach picosecond timescales at low computational cost. This enables applications such as the study of the optical properties of nanomaterials or biomolecules~\cite{Douglas-Gallardo2016a,MansillaWettstein2016a}, the dynamics of optically excited vibrational modes in nanostructures~\cite{Bonafe2017a} and other challenging systems. Real-time time-dependent density-functional tight-binding (TD-DFTB) is a suitable formalism for these cases which has been widely used to calculate quantum-mechanically ground state static and dynamic properties of large atomistic systems~\cite{Oviedo2010,Negre:2012fm,Douglas-Gallardo2016a,MansillaWettstein2016a,Medrano2016a}.

Here, we present the theoretical grounds and relevant computational details of a real-time Ehrenfest TD-DFTB implementation\footnote{Available at \texttt{https://github.com/dftbplus/dftbplus}} in the DFTB+ code~\cite{Hourahine2020}, explained in section \ref{sec:theory}. The possibility to calculate the optical properties of periodic systems that this implementation allows is discussed in section \ref{sec:gnr} for graphene nanoribbons. The DFTB+ code not only provides an efficient and parallel way to calculate the ground state of molecular and periodic systems, i.e.\ the initial state for the dynamics, but also includes several corrections to the density functional tight-binding (DFTB) Hamiltonian. These include the approximate self-interaction corrections (``LDA+U'' and ``pseudo SIC'' methods~\cite{hourahine2007self}) and 3rd order DFTB~\cite{gaus2011dftb3} (or ``DFTB3'') , which can be used to improve the modeling of time-dependent properties in some systems, as it is shown in section \ref{sec:dftbu}. In section \ref{sec:nanodiamond} we discuss the application of this method to the calculation of the effect of the nuclei dynamics in the charge transfer profile in a donor-acceptor interface. The present implementation, within the limitations of Ehrenfest dynamics, is also able to simulate time-resolved transient absorption spectra and impulsive vibrational spectra, as far as we are aware, for the first time accounting for real-time propagation of the nuclei in an atomistic trajectory \revone{of several hundreds of femtoseconds}~\cite{Bonafe2018}. This development is presented in section \ref{sec:tas} and is applied to study the ultra-fast dynamical features of on- and off-resonance excitation of zinc-tetraphenylporphyrin (ZnTPP) in section \ref{sec:znttp}. Finally, general conclusions and future prospects for this method are discussed in section \ref{sec:conclusions}.

\section{Theory and implementation details}
\label{sec:theory}

\subsection{Theoretical background}

The Ehrenfest propagation scheme reported hereafter is based on the DFTB formalism. The DFTB method has been explained in detail in several references~\cite{Elstner1998,Frauenheim2002b,Koskinen2009a}. The following description summarizes the approximations of the method with respect to DFT. For simplicity we will consider the spin-unpolarized expressions, but the development has been done accounting for systems with collinear spin polarization. Starting from the energy expression as a functional of the density in DFT, $E[n]$, the second order DFTB method is obtained by expanding the energy around a reference density $n_0$ up to second order in density fluctuations $\delta n$~\cite{Frauenheim2002b}:
\begin{equation}
\begin{split}
 E[n] \approx&  \sum_i f_i \langle \psi_i|-\frac{1}{2} \nabla^2 + V_{ext} + V_H[n_0] + V_{xc}[n_0] |\psi_i \rangle \\
  &+ \frac{1}{2} \int \int d\br d\br' \; \left(\frac{\delta^2E_{xc}[n_o]}{\delta n \delta n'} + \frac{1}{|\br-\br'|} \right) \delta n \delta n' \\
  &-\frac{1}{2} \int V_H[n_0](\br)n_0(\br) + E_{xc}[n_0] + E_{AA} \\
  &- \int d\br \; V_{xc} [n_0](\br)n_0(\br),
\end{split}
\label{eq:endeltan}
\end{equation}
where $|\psi_i \rangle$ are the molecular orbitals, $f_i$ is the occupation of the $\ket{\psi_i}$ state, $V_{ext}$, $V_H$ and $V_{xc}$ are the external, Hartree and exchange correlation potentials, $E_{xc}$ and $E_{AA}$ are the exchange-correlation and inter-nuclear repulsion energies and $\br$ is the electronic position. A pseudo-atomic orbital basis, $\{\ket{\phi_\mu}\}$ is considered and, using the two-center approximation for off-diagonal interactions, the matrix elements of the non self-consistent Hamiltonian $H_0$ can be obtained and parameterized as a function of the inter-atomic distance. This leads to a simple expression for the first term of eq.~\eqref{eq:endeltan} (the non-SCC \revone{single particle} energy):
\begin{align}
E_{band} = \sum_i f_i \langle \psi_i | H^0 | \psi_i \rangle = \Tr \{\rho H^0\},
\end{align}
where we have introduced the one-body reduced density matrix, $\rho$ and $\Tr$ represents the trace. The atomic contributions to the charge density are approximated by monopolar Mulliken charge fluctuations at the atoms. Assuming a spherical symmetry of the density fluctuation, the Coulomb integrals in the second term of eq.~\eqref{eq:endeltan} can be solved and then the coulombic contribution is written as sums of products of atomic quantities:
\begin{align}
E_{coul} &= \frac{1}{2} \sum_{AB} \gamma_{AB} \Delta q_A \Delta q_B, \label{eq:coul} \\
\gamma_{AB} &= \left\{ \begin{array}{lr}
                       U_A, & A=B \\
                       \frac{1}{R_{AB}}-\zeta(R_{AB}), & A\neq B
                      \end{array} \right.,
\end{align}
where the parameters needed for the onsite terms (the Hubbard parameters $U_A$) and the short range terms $\zeta(R_{AB})$ are obtained from \textit{ab initio} calculations. The short range term damps the Coulomb expression for short ranges and ensures the right limit at $R_{AB} \to 0$. The Mulliken populations are calculated as:
\begin{equation}
 q_A = \sum_{\nu \in A} \sum_{\mu} \rho_{\nu \mu} \bra{\phi_\mu} \phi_\nu\rangle = \Tr_A \{ \rho S \},
 \label{eq:qmulik}
\end{equation}
where $\Tr_A$ indicates the trace of over the orbitals centered on atom $A$, and the overlap matrix $S$ depends on the inter-atomic distance and is parameterized for each pair of orbitals. \revone{Beyond-monopole corrections have been previuosly developed and discussed for Slater-type-orbital-based DFTB \cite{Bodrog2012,Dominguez2013a,Dominguez2015a} and also for similar methods like Gaussian tight binding \cite{Boleininger2016, Boleininger2017b}, which adds further contributions to the Coulomb terms of eq. \ref{eq:coul} and to the dipole matrix elements, which modifies directly the oscillator strengths. This is important to accurately describe the absorption spectra of molecules with low-lying $\sigma \to \pi^*$ and $n \to \pi^*$ excitations \cite{Dominguez2013a}. However, for most of the large molecular systems studied with the real-time Ehrenfest TDDFTB formalism, accurate results can be obtained in the monopole approximation at a very low computational cost\cite{Niehaus2009a,Oviedo2010,Oviedo:2011jt,Negre:2012fm,Medrano2016a,MansillaWettstein2016a,Bonafe2017a}. Hence, the monopole approximation has been used throughout this work.}

The last four terms of eq.~\eqref{eq:endeltan} do not depend on the density fluctuation and hence can be collected into one single term, called the repulsion energy, $E_{rep}$, which \revone{are} decomposed into pair-wise atomic contributions \revone{(neglecting single site terms, these being irrelevant for cohesive properties and relative energies)}, $V^{rep}_{AB}$, that are also parameterized as a function of the inter-atomic distance. Hence, evaluation of the energy expression does not involve the computation of molecular integrals, which makes this approach extremely fast. The necessary parameters $H_0$ and $S$ are calculated in a minimal basis of valence orbitals for each pair of orbitals between each pair of atomic species at different inter-atomic distances. The repulsive terms, $V^{rep}_{AB}$, are fitted minimizing DFTB and DFT energy and force differences for a set of systems that span a representative range between the pairs of atomic species $A$ and $B$. These are the only semi-empirical parameters in the DFTB method. Several parameters sets including the ones used in the present work are available for download at \url{http://www.dftb.org}.

By minimizing the variation of the energy with respect to the expansion coefficients of the orthogonal molecular orbitals in the pseudo-atomic basic, $\{C_{i\mu}\}$, given by $\ket{\psi_i} = \sum_\nu C_{i\mu} \ket{\phi_\mu}$, an expression for the DFTB Hamiltonian can be obtained:
\begin{equation}
 H_{\mu \nu} = H^0_{\mu \nu} + \frac{1}{2} S_{\mu \nu} \sum_C (\gamma_{AC} + \gamma_{BC}) \Delta q_C, \;\;\; \mu \in A, \nu \in B.
 \label{eq:hscc}
\end{equation}

When an external potential is applied the Hamiltonian matrix changes to:
\begin{equation}
 H_{\mu \nu} \rightarrow H_{\mu \nu} + \frac{1}{2} S_{\mu \nu} (V^{ext}_A + V^{ext}_B), \;\;\; \mu \in A, \nu \in B.
 \label{eq:hsccvext}
\end{equation}
\revone{The eigenvalue equation derived from the SCC Hamiltonian defined in eq.~\eqref{eq:hsccvext}} can be solved self-consistently to obtain the ground state (GS) of the system, by starting from a guessed charge distribution followed by consecutive cycles of diagonalization and Hamiltonian building until self-consistency. This method is known as self-consistent charge DFTB (SCC-DFTB).

Derivation of the equations of motion for wavefunction coefficients and nuclear positions according to the Ehrenfest method for tight-binding schemes have already been published~\cite{Todorov:2001ex,Niehaus:2005da}, and are well known from TD-DFT, where the time-dependent Kohn-Sham equations can be obtained from the Runge-Gross theorem~\cite{Runge1984}. The equation of motion (EOM) of the coefficients $C_{i\mu}$ reads:
\begin{align}
\dot{C}_{i \mu} = -\im \sum_{\kappa \nu} \invs_{\kappa \mu} \left( H_{\kappa \nu} - \im \langle \phi_\kappa|\dot{\phi_\nu}\rangle \right)C_{i \nu}.
\label{eq:eomcoeff}
\end{align}
All the matrix elements on the right-hand side \revone{have already been introduced throughout this text}, except for the non-adiabatic coupling (NAC) matrix elements:
\begin{equation}
D_{\kappa \nu} = \langle \phi_\kappa|\dot{\phi_\nu}\rangle.
\end{equation}
As the pseudo-atomic basis set is centered on the atomic nuclei, these matrix elements are non-zero only when the nuclei move. The EOM of the density matrix can be obtained, as shown in the appendix~\ref{sec:derivationeom}, from eq.~\eqref{eq:eomcoeff}:
\begin{equation}
 \dot{\rho} = -\im (\invs H \rho - \rho H \invs)- (\invs D \rho + \rho D^\dagger \invs).
\label{eq:eomrho}
\end{equation}
The first term of the right-hand side of this equation is the electronic EOM known as the Liouville-von~Neumann equation~\cite{Breuer2010}, a generalized commutator in a non-orthogonal basis, and it has been implemented and extensively used for the calculation of optical properties of materials and time-dependent photo-induced phenomena keeping the nuclei ``clamped''~\cite{Oviedo2010, Oviedo:2011jt,Negre:2012fm,Oviedo:2012ct, MansillaWettstein2016a, Douglas-Gallardo2016a,Medrano2016a}. The second term gives rise to the non-adiabatic interactions, allowing energy exchange between electrons and nuclei at the Ehrenfest level.

As the basis orbitals' time evolution is tied to the nuclei on which they are centered, the NAC elements can be computed from the chain rule. By making explicit the nuclear index where each orbital is centered, the $D$ matrix elements are then:
\begin{equation}
 D_{\mu \nu} = \langle \phi_{\mu} | \dot{\phi}_{\nu} \rangle = \langle
 \phi_{\mu} | \nabla_B \phi_{\nu} \rangle \cdot \dot{\bR}_B, \; \mu \in A,
 \nu \in B,
\end{equation}
where $\nabla_B$ is the gradient applied with respect to the coordinates of atom $B$ and $\dot{\bR}$ is the velocity. To calculate $ \langle \phi_{\mu} | \nabla_B \phi_{\nu} \rangle$ we use the fact that $S_{\mu \nu}$ is only a function of $\bR_A - \bR_B$, and hence for orbitals sitting at different atoms $A \neq B$:
\begin{equation}
 \nabla_B S_{\mu \nu} = \langle \nabla_B \phi_{\mu} | \phi_{\nu} \rangle + \langle \phi_{\mu} | \nabla_B \phi_{\nu} \rangle = \langle \phi_\mu | \nabla_B \phi_\nu \rangle,
 \label{eq:gradover}
\end{equation}
since $\nabla_B \phi_{\mu} = 0$. This result is equivalent to the one derived in ref.~\citenum{Todorov:2001ex}. \revtwo{Formally,} for orbitals sitting on the same atom (onsite blocks of $D$), the limit of the gradient in eq.~\eqref{eq:gradover} as the two atoms approach each other must be taken:
\begin{equation}
\langle \phi_{\mu_{A}} | \nabla_A \phi_{\nu_A} \rangle = \lim_{|\bR_{A'} - \bR_A| \rightarrow 0}  \nabla_A  S_{\mu_{A'} \nu_A}.
\label{eq:onsite}
\end{equation}
\revtwo{Considering that the $D$ matrix elements can be related to the linear momentum matrix elements~\cite{Todorov:2001ex}, the onsite elements can be regarded as the small internuclear distance limit of an atomic scattering problem. This way it can be argued that the onsite blocks can been set to zero~\cite{Kurpick1995}. This have been used and validated in previous Ehrenfest implementations~\cite{Niehaus:2005da}, and in this implementation by comparison with reported light-induced impulsive vibrations \cite{Bonafe2017a,Bonafe2018}. The following general expression for the NAC matrix elements can then be written:}
\begin{equation}
 D_{\mu \nu} = \dot{\bR}_B \cdot \nabla_B S_{\mu \nu} (1 - \delta_{AB}).
 \label{eq:nac}
\end{equation}

From eq.~\eqref{eq:nac}, two conclusions can be drawn:
\begin{enumerate}[(a)]
\item For fixed nuclei, $D=0$ and the electronic EOM, the first term in eq.~\eqref{eq:eomrho}, is recovered.
\item It can be verified that $D^\dagger \neq -D$. If they were equal, this would allow the effective Hamiltonian $\tilde{H}$ to be hermitian (see appendix \ref{sec:derivationeom}), making the evolution unitary, with no change in the eigenvalues of the density matrix. This only happens when all velocities are equal, which is the case for center of mass motion only without rotation. Hence, unitary evolution is recovered when there is no relative motion of the atoms.
\end{enumerate}

The EOM for the nuclei has also been derived in previous works, starting from the Lagrangian of the system~\cite{Todorov:2001ex,Niehaus:2005da}. The Euler-Lagrange equations for the coefficients lead to the EOM for the coefficients, eq.~\eqref{eq:eomcoeff}, while differentiating the Lagrangian with respect to the nuclear coordinates leads to the expression for the Ehrenfest force. Rewritten here in terms of the density matrix this is:
\begin{align}
  \bF_A =& -\Tr \left\{ \rho \left( \nabla_A H^0 + \nabla_A S \sum_B \gamma_{AB} \Delta q_B + (\nabla_A S) \invs H +  H \invs (\nabla_A S) \right) \right\} \nonumber \\
 &- \im \; \Tr \left\{ \rho (\nabla_A S) \invs D + \mathrm{h.c.} \right\} + \im \; \sum_{\mu \nu} \left\{ \rho_{\nu \mu} \langle \nabla_A \phi_{\mu} | \nabla_B \phi_{\nu} \rangle \cdot \dot{\bR}_B + \mathrm{h.c.} \right\} \nonumber \\
  &- \Delta q_A \sum_{B} \nabla_A \gamma_{AB} \Delta q_B - \nabla_A E_{rep} - \Delta q_A \bE(t),  \;\;\; \mu \in A,
 \nu \in B,
  \label{eq:forces}
\end{align}
where $\bE(t)$ is the time-dependent external field, $\mathrm{h.c.}$ is the hermitian conjugate and the second and third terms depend on the  nuclear velocities, which cancel each other out for a complete basis~\cite{Niehaus:2005da}. These terms are necessary for momentum conservation but not for energy conservation, since the latter emerges from the lack of explicit time dependence of the Lagrangian, nor do they produce net work~\cite{Todorov:2001ex}. Total charge conservation, however, is given by the velocity-dependent terms in the electronic EOM, which are indeed included (namely, the NAC terms). In the present implementation the velocity-dependent terms of the nuclear forces are ignored, since for all the cases of interest considered here these terms are negligible due to the small nuclear velocities involved. Moreover, the fact that the forces depend only on the electronic state and the nuclear positions enables more efficient methods for propagating the nuclei. Otherwise, other methods typically used for dynamics with dissipative forces must be used, where the forces and velocities must be solved self-consistently at each time step.

To drive the system, a time-dependent external field can be added to the DFTB Hamiltonian according to eq.~\eqref{eq:hsccvext}, where the external potential can be calculated under the electric dipole approximation: $$V^{ext}_A(t) = -\bmu_A(t) \cdot \bE(t) = \Delta q_A(t) \bR_A(t) \cdot \bE(t).$$

The limitations of the Ehrenfest method, such as its inability to reproduce Joule heating on the nanoscale~\cite{Horsfield2004c} or the photoisomerization of retinal models~\cite{Niehaus:2005da}, are well known. The lack of nuclear fluctuations suppresses heat dissipation to the nuclei from excited electrons~\cite{Horsfield2004c}. To observe inelastic dissipation, other methods such as lowest-order electron-phonon scattering theory are needed, where the electron-phonon interaction is treated by first-order perturbation theory~\cite{Montgomery2002}. This theory however fails in the strong electron-phonon coupling regime and other methods such as the self-consistent Born approximation~\cite{Haug2008} or the coupled electron ion dynamics (CEID) method~\cite{McEniry2008} are necessary.

On the other hand, the method is fully quantum coherent, since the non-diagonal
elements of the DM (coherences) are included in the dynamics. Besides, it has been proven that the nuclear motion, through its effect in the Hamiltonian, can induce electronic transitions, mainly in metals, and can lead to \revone{an electronic state that can be characterised by a temperature ~\cite{Lin2009d}. However, although the electrons achieve a Fermi-Dirac distribution, its temperature is different from that of the nuclei, and hence cannot be regarded as an equilibration process.}

\subsection{Ground state absorption spectra}
\label{sec:gsspec}

The simulation protocol to compute the GS absorption has been derived two decades ago and has been used in several research works by different groups. This short summary highlights the main concepts and will be useful in section \ref{sec:tas}. First, the ground state DM $\rho_0$ is \textit{kicked} by a Dirac-delta electric field pulse $\bE_{kick} (t) =  \Ef_0 \delta(t) \be$ polarized in direction $\be \in \{\hat{i},\hat{j},\hat{k}\}$, exciting all dipole allowed transitions \cite{Yabana1996b}. For a non-orthogonal basis set, this DM immediately after a Dirac-delta perturbation is calculated according to eq. \ref{eq:rhokick} \cite{Negre2010a}:
\begin{equation}
 \rho(t=0^+) = \frac{1}{2}\left( e^{\frac{i}{\hbar}\hat{V}} \rho_0 S e^{-\frac{i}{\hbar}\hat{V}} S^{-1} + S^{-1} e^{\frac{i}{\hbar}\hat{V}} S \rho_0 e^{-\frac{i}{\hbar}\hat{V}} \right)
 \label{eq:rhokick}
\end{equation}
where $\hat{V} = -\hat{\bmu} \cdot \bE_{kick}$. The system then evolves freely and the time-dependent dipole moment of the system contains the information about the excited frequencies and oscillator strengths. By Fourier transforming the dipole moment signal along each cartesian direction after exciting in any other cartesian direction, the polarizability tensor (Fourier transform of the response function) for all angular frequencies $\omega$ can be found \cite{Castro2006a}:
\begin{equation}
 \bpol(\omega) = \frac{\bmu(\omega)-\bmu_0}{\Ef_0},
 \label{eq:alpha}
\end{equation}
where $\Ef_0$ is the Dirac-delta \revone{(scalar)} field strength. The absorption cross-section for randomly oriented molecules can be calculated as \cite{Onida2002}:
\begin{equation}
 \sigma(\omega) = \frac{4\pi\omega}{c}\mathrm{Im}(\bar{\alpha}(\omega)),
 \label{eq:sigma}
\end{equation}
where $\bar{\alpha}(\omega) = \frac{1}{3}\Tr[\bpol(\omega))]$. This method has
been used by some of us to calculate absorption spectra of molecular systems~\cite{Oviedo2010,Oviedo:2011jt,Medrano2016a}, semiconductor~\cite{Negre:2012fm,Oviedo:2012ct,Fuertes2013,MansillaWettstein2018} and metallic nanoparticles~\cite{Negre2013a,Douglas-Gallardo2016a}, as well as carbon-based nanostructures~\cite{MansillaWettstein2016a}.

\subsection{Implementation}

The implementation takes advantage of the GS and structural relaxation currently implemented in DFTB+. The time evolution is achieved by time discretization and numerical propagation of the DM and nuclear coordinates in the time grid. The integration of the density matrix is carried out using the Leapfrog technique, a second-order method, given by the following update equation:
\begin{equation}
\rho_{i+1} = \rho_{i-1} + 2\Delta t \dot{\rho}_i,
\label{eq:leapfrog}
\end{equation}
where $\dot{\rho}_i$ is calculated using eq.~\eqref{eq:eomrho}. The integration step that ensures stable dynamics for most systems is between 1-5 as. As it is a two-step method, it must be initialized using a single-step method to compute the initial states for propagation; here, the Euler method is used. The bottleneck of this method is given by the calculation of $\dot{\rho}$ which can be done in three matrix-matrix multiplications, computed here using optimized BLAS subroutines and with a maximum scaling with the size of the system (number of basis orbitals $N$) of $\mathcal{O}(N^3)$. This method has the advantage over other choices such as the Crank-Nicholson or Magnus expansion that it only needs the Hamiltonian at the present time-step to compute the DM at the following time-step, so no extrapolation or predictor-corrector scheme is necessary. Moreover, it exactly conserves the number of electrons and is time-reversible.

The NAC matrix elements are calculated using eq.~\eqref{eq:nac} where the gradients of $S$ are approximated by finite differences. The derivative with respect to atom $B$ in the direction $\alpha$ is given by:
\begin{equation}
\frac{\partial}{\partial R_{B_\alpha}}S = \frac{1}{2\Delta R}(S(R_{AB_\alpha}+\Delta R)-S(R_{AB_\alpha}-\Delta R)).
\label{eq:overprime}
\end{equation}
Here $\bR_{AB} = \bR_B - \bR_A$ and $\Delta R = \sqrt[4]{\epsilon}$ where
$\epsilon$ is the double precision machine epsilon. $\Delta R$ is
  chosen since the rounding error for a step of size $h$ is of the order of
  $\epsilon/h^2$ and the central difference derivative formula has a truncation
  error of $\mathcal{O}(h^2)$.

The forces (eq.~\eqref{eq:forces}) are computed using the sparse matrix subroutines included in DFTB+~\cite{Aradi2007b}, using numerical approximations for $\nabla_A H$ and $\nabla_A S$, and the analytical expressions for $\nabla_A \gamma_{AB}$ and $\nabla_A E_{rep}$. The propagation of the nuclear coordinates uses the well-known velocity Verlet method~\cite{Swope1982}.

\section{Novel simulations of optical properties}

\subsection{Optical properties of graphene nanoribbons}
\label{sec:gnr}

After the experimental synthesis of graphene, by Geim and Novoselov in 2004 \cite{Novoselov2004}, it has been considered a promising material for future electronics. Due to its extremely high charge-carrier mobilities \cite{Novoselov2004,Chen2008}, graphene-based transistors have been developed rapidly and considered as a choice for post-silicon electronics \cite{Schwierz2010}. However, the lack of a bandgap in graphene hinders its application in field-effect transistors devices \cite{Schwierz2010}. In this sense, flat one-dimensional sp$^2$ carbon based materials, known as graphene nanoribbons (GNRs), are proposed as the building blocks for the next-generation in this field.
Unlike infinite graphene, GNRs show non-zero bandgaps that strongly depend on the ribbon width \cite{Son2006,Son2006_2,Wang2016}. Particularly, armchair GNRs (aGNRs) are the most promising candidates in this area because of the high tunability of their electronic properties \cite{Wang2016}.
In this context, the study of the optoelectronic properties of these kind of structures is necessary in order to drive the development of new carbon-based nanoelectronic technologies.

\begin{figure*}[ht]
    \centering
    \subfloat[]{%
     \includegraphics[width=0.9\textwidth]{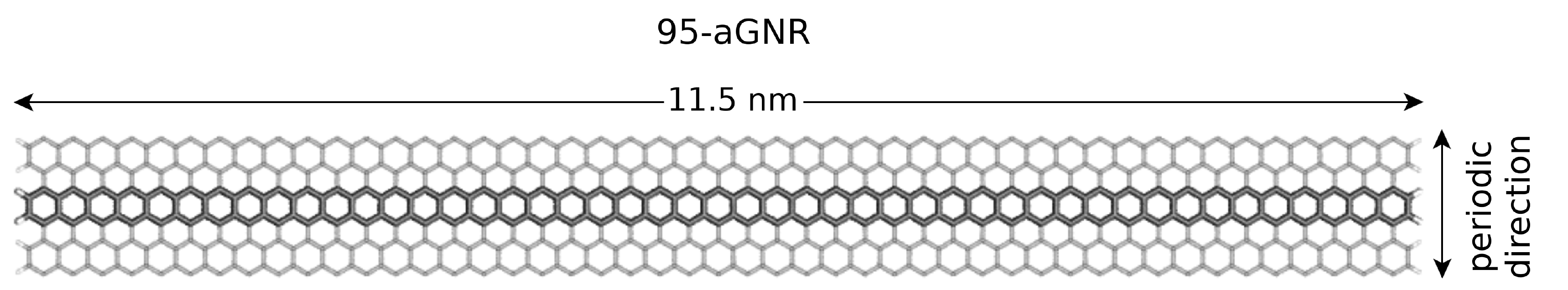}
     \label{ribbon}
    }\\
    \subfloat[]{%
     \includegraphics[height=0.45\textwidth]{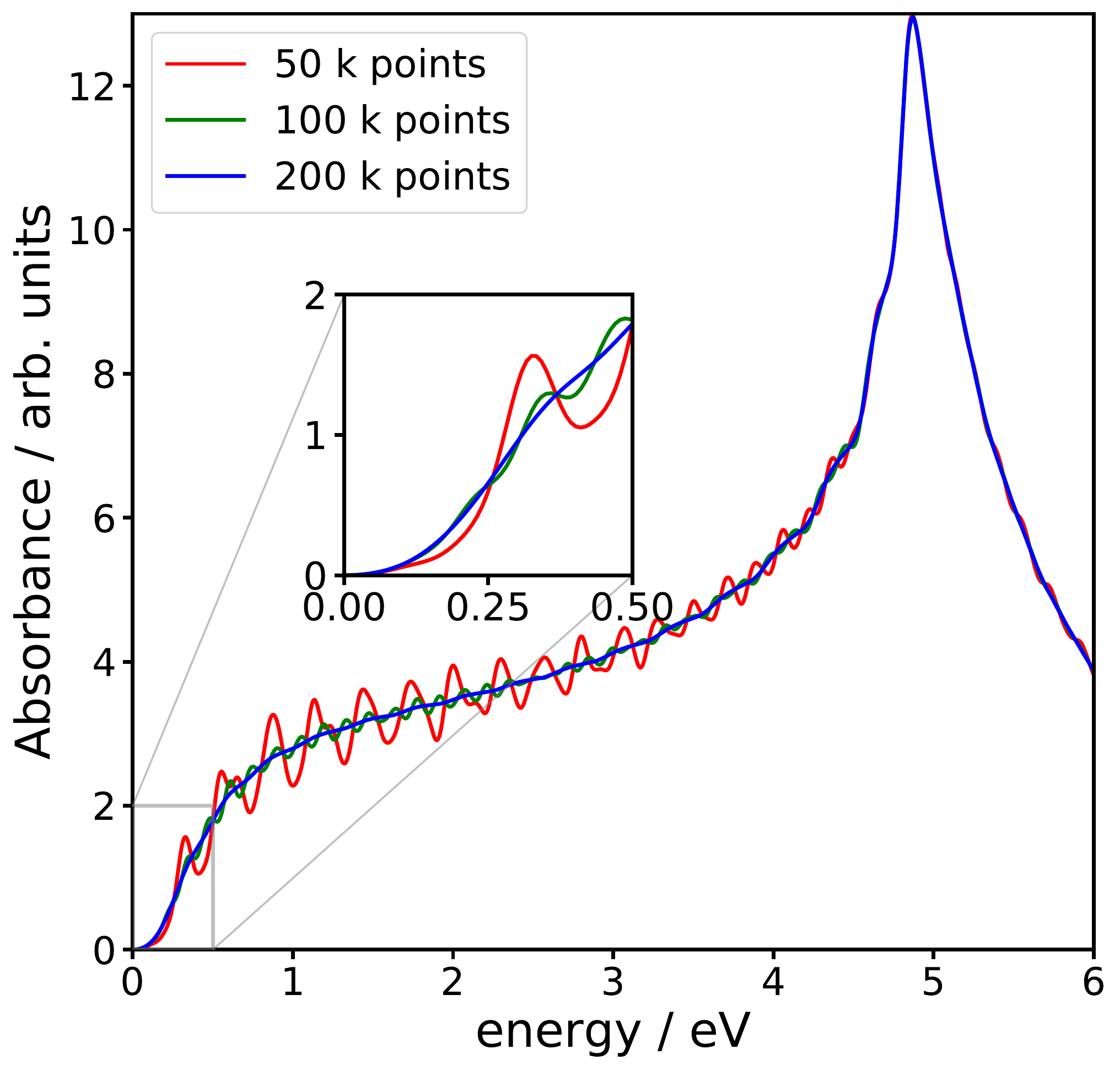}
     \label{ribbon1}
    }
    \subfloat[]{%
     \includegraphics[height=0.45\textwidth]{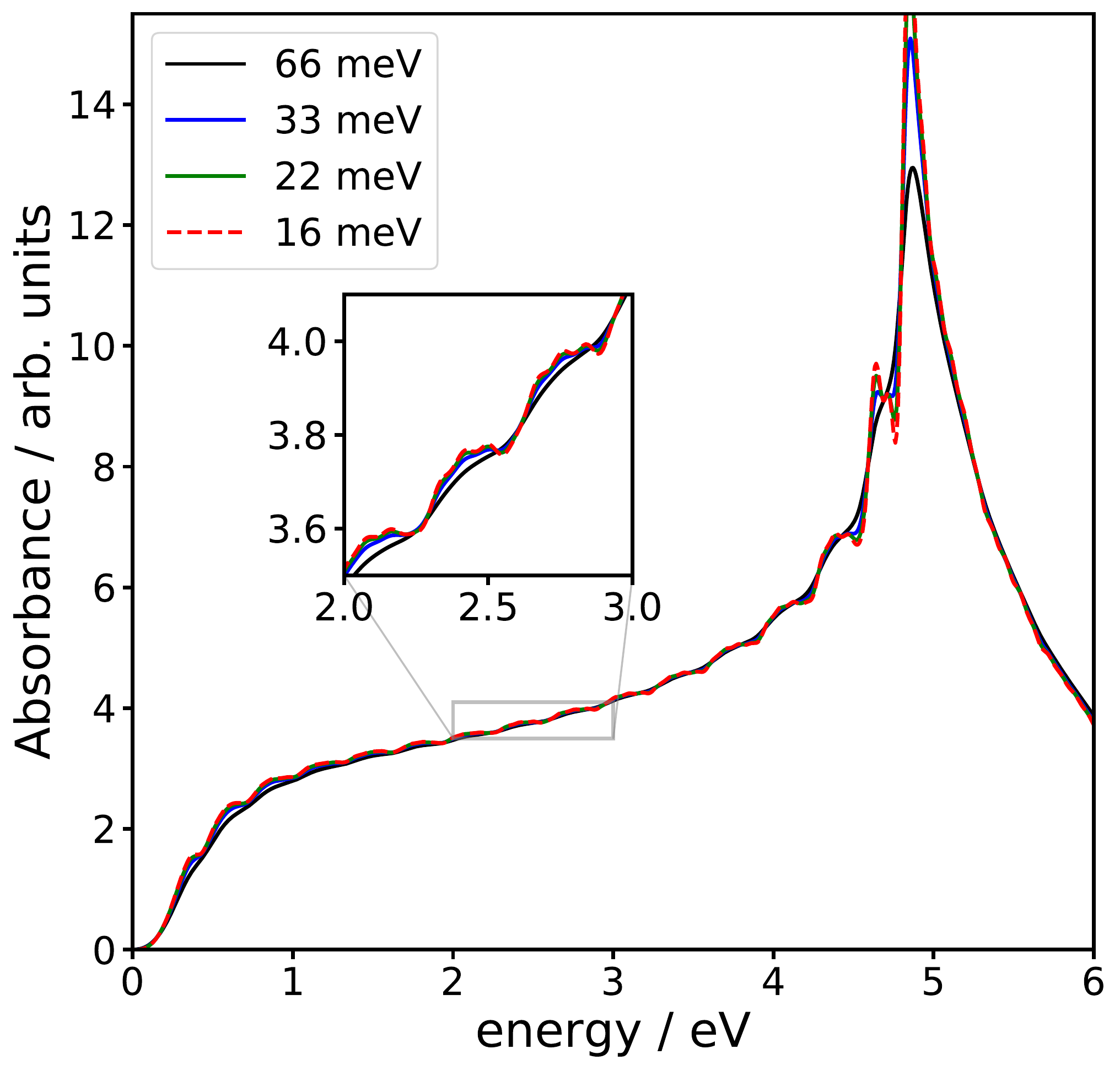}
     \label{ribbon2}
    }
    \caption{(a) Unit-cell representation for the 95-aGNR studied. (b) Electronic spectra in the direction normal to the ribbon axis for the semiconducting 95-aGNR of the $3p+2$ family using different numbers of $\mathbf{k}$ points to sample the periodic direction of the first Brillouin zone. A broadening of 66 meV, corresponding to an exponential damping decay time of 10 fs, was used in all cases. (c) Same as (b) for different broadenings. 200 $\mathbf{k}$-point samples within the first Brillouin zone in the periodic direction where used in all cases.}
    \label{fig:nanoribbon}
\end{figure*}

The absorption spectrum of a 95 atom wide arm-chair graphene nanorribon (95-aGNR, see figure \ref{ribbon}) was calculated as described in section \ref{sec:gsspec} as an example of the possibility to calculate spectra of lower-dimensional structures. \revtwo{The mio-1-1 Slater-Koster parameter set was used \cite{PhysRevB.58.7260}}. The absorption spectrum for light polarized in the plane of the ribbon and normal to the periodic direction is shown in figure \ref{fig:nanoribbon} for different $\mathbf{k}$-point samplings and signal damping times. The calculated absorption shows an optical gap of about 10 meV, obtained by linear extrapolation of the absorption close to the gap. This small gap is expected for an armchair nanorribon of the $3p+2$ family \cite{Young2006}. The absorption raises rapidly with energy between 0 to 1 eV and then slowly to about 4 eV. A sharp peak is present in the absorbance at 5 eV which is about five times higher than the absorption at 2 eV. These features are in good agreement with existing results in the literature \cite{Lian2018} obtained using a very similar method to the one present in this work but based on time dependent DFT. The spectra shown in figure \ref{fig:nanoribbon} are also in good agreement to the calculations presented by Yang {\em et al.} \cite{Yang2009} for graphene not including excitonic effects. When compared to the results in ref. \citenum{Yang2009} that include excitonic effects, the peak at 5 eV in figure \ref{fig:nanoribbon} is at higher energies but lower than those shown by Lian {\em et al.} in ref.  \citenum{Lian2018}. The fact that results using TD-DFTB improve upon those of TD-DFT has been observed by some of us before \cite{Oviedo:2011jt} and clearly corresponds to a compensation of errors arising from the model. 

% \begin{figure*}[ht]
%     \centering
%     \subfloat[]{%
%      \includegraphics[width=0.45\textwidth]{Ribbon-figs/nribbon-k-convergence.pdf}
%      \label{ribbon1}
%     }
%     \subfloat[]{%
%      \includegraphics[width=0.45\textwidth]{Ribbon-figs/nribbon-tau-convergence.pdf}
%      \label{ribbon2}
%     }
%     \caption{(a) Electronic spectra in the direction normal to the ribbon axis for the semiconducting 95-aGNR of the $3p+2$ family using different numbers of $\mathbf{k}$ points to sample the periodic direction of the first Brillouin zone. A broadening of 66 meV, corresponding to an exponential damping decay time of 10 fs, was used in all cases. (b) Same as (a) for different broadenings. 200 $\mathbf{k}$-point samples within the first Brillouin zone in the periodic direction where used in all cases.}
%     \label{fig:nanoribbon}
% \end{figure*}

Regarding the large number of $\mathbf{k}$-point samples that are required to converge absorption spectra, as can be seen from figure \ref{ribbon1}, a large density of points is required. This is to be expected as a similar density of sampling would be needed to obtain a converged density of states. Once the spectrum is converged with respect to sampling of the Brillouin zone, as depicted in figure \ref{ribbon2}, the spectrum is independent of the broadening or damping used before the Fourier transform of the time dependent dipole moment.

\subsection{DFTB+U spectra for Ni(CO)$_4$}
\label{sec:dftbu}

\begin{figure}
  \includegraphics[width=.5\textwidth]{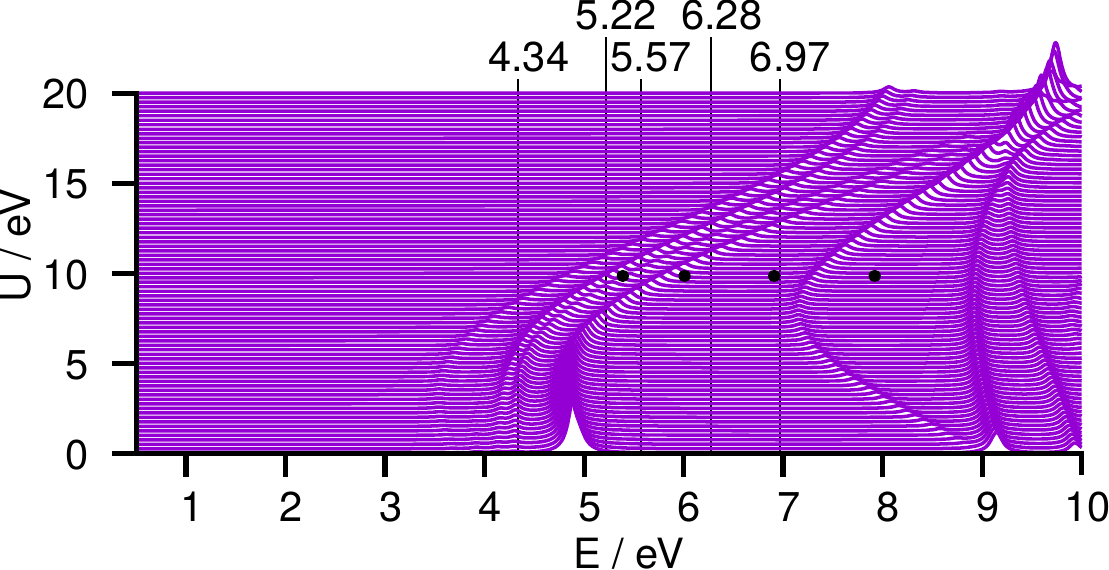}
  \caption{\label{fig:plusU} Electronic spectra of \cf{Ni(CO)4} for ground state
    relaxed DFTB+U geometries. The marked circles correspond to TD-GGA+U
    results~\cite{PhysRevB.99.165120}, with the CASPT2 results of
    ref.~\citenum{pierloot1996optical} marked along the top of the figure.}
\end{figure}

For systems where there is a strong interaction between localised electrons,
such as transition metal or lanthanide compounds and solids, a simple mean-field
correction for electron correlation is provided by the LDA+U family of
methods~\cite{Anisimov_1997}. The rotationally invariant~\cite{PhysRevB.57.1505}
form of LDA+U can be written in terms of several choices of local projections of
the density matrix~\cite{PhysRevB.73.045110}. Likewise, the double-counting
between the Hubbard-model and the density functional mean-field functional take
several limiting cases~\cite{PhysRevB.67.153106}. In the DFTB+ code the
dual population projected form of the fully-localised limit (FLL) functional has
been available for ground state calculations for over a
decade~\cite{doi:10.1021/jp070173b}. The additions to the DFTB energy and Hamiltonian take the form:
\revone{
\begin{align}
\Delta E^{FLL} &= - \sum_A \sum_{l} \sum_{\mu\nu} \frac{U^{\text{eff}}_l}{2} \left(n_{\mu\nu}^2 - n_{\mu\mu}\right)\\
\Delta V^{FLL}_{\mu\nu} &= - U^{\text{eff}}_l S_{\mu\nu} \left( n_{\mu\nu} - \frac{1}{2}\delta_{\mu\nu} \right)\\
n_{\mu\nu} &= \frac{1}{2} \sum_{\mu\nu} \sum_\kappa \left( \rho_{\mu\kappa}S_{\kappa\nu} +
  S_{\mu\kappa} \rho_{\kappa\nu} \right),
\end{align}
where $\{\mu,\nu\}\in l \in A$, i.e. the orbitals within the angular shell $l$ on atom $A$, but $\kappa$ is summed over all orbitals for which the overlap to that atomic block is non-zero.
}
%\begin{align}
%  \Delta E^{FLL} &= - \sum_A \sum_{l \in %A}\frac{U^{\text{eff}}_l}{2} \left(n_l^2 - n_l\right)\\
%  %
%  \Delta V^{FLL}_l &= - U^{\text{eff}} S_l \left( n_l - %\frac{1}{2} \right)\\
%  %
%  n_l &= \frac{1}{2} \sum_{\mu \in l} \sum_{\nu \in l} %\sum_\kappa \left( \rho_{\mu\kappa}S_{\kappa\nu} +
%  S_{\mu\kappa} \rho_{\kappa\nu} \right),
%\end{align}
%where $l$ is a block-wise population of the orbitals in that shell on atom $A$
%and $\mu$ and $\nu$ refer to particular orbitals in that block. 

The choice of the \emph{effective} Hubbard-U parameter, $U^{\text{eff}}$, is a
contentious issue with a number of recipes in the literature, ranging from an
empirical choice to reproduce experiment through to recipes for screened
self-consistent evaluation. Figure~\ref{fig:plusU} shows the excitation energies
of gas phase \cf{Ni(CO)4} from one self-consistent GGA+U
approach~\cite{PhysRevB.99.165120}, marked as circles for the value of $U^{\text{eff}}$
and the resulting optically active transition energies. This figure also shows
the results of correlated wave-function based
calculations~\cite{pierloot1996optical}. These are compared against DFTB+U (the
purple traces) for a range of $U^{\text{eff}}$ values from uncorrelated ($U^{\text{eff}} = 0$)
to approximately twice the unscreened atomic values for \cf{Ni^{3d}} at the GGA
level (20~eV). The trans3d~\cite{doi:10.1021/ct600312f} and
mio~\cite{PhysRevB.58.7260} Slater-Koster parameter sets were used, where the
shell-resolved form of the DFTB Hamiltonian was applied (as a
result the mio \cf{C} and \cf{O} parameters were edited to set atomic Hubbard-U
values to their highest atomic occupied values, correcting the known error in
these parameters when used for shell resolved calculations). The DFTB+U
absorption spectra were obtained by optimizing the GS geometry at each
value of $U$ and using this as the source of the initial density matrix. The
absorption spectra were then calculated for a fixed geometry by applying the
approach of section~\ref{sec:gsspec} using the DFTB+U Hamiltonian to propagate the
density matrix.

The closest match to the CASPT2 results is obtained for an effective $U_{3d}$
value of $\sim$8~eV, where the DFTB+U features most closely approach the
excitation energies of this method. This is slightly lower than the 9.85~eV
effective $U$ determined self-consistently for GGA. This is primarily due to the
lower- and particularly the upper-most absorption features in the DFTB+U
spectra, which most closely approach the CASPT2 features at a lower value of
$U$. The three central features in the group instead best match these results
for a value of $U$ slightly above 10~eV, leading to the effective value for best
spectroscopic properties on average.
% Addition to paper from BH for referee 2 comment:
\revtwo{Unfortunately it is not at present possible to verify the symmetry assignments of the modes with real time propagation methods (since the density matrix transforms with the full symmetry of the Hamiltonian, apart from the spin and the effects of spin contamination~\cite{doi:10.1002/qua.24309}). But the recent work of Oppenheim {\it et al.}~\cite{doi:10.1021/acs.inorgchem.9b02135} demonstrates that the symmetries of the excitations in a related nickel compound are correctly reproduced for a range of density-matrix functionals.}

Pseudo-SIC~\cite{PhysRevB.67.125109} instead approximately removes the local
part of the self-interaction error, but only affects the occupied states of the
system, unlike FLL which is similar to a scissor operation on the affected
shell. Again, this can be expressed in the local block populations:
\begin{align}
  \Delta E^{FLL} &= - \sum_A \sum_{l \in A}\frac{\tilde{U}^{\text{eff}}_l}{2} n_l^2\\
  \Delta V^{FLL}_l &= - \tilde{U}^{\text{eff}} S_l n_l
\end{align}
where $\tilde{U}$ is approximately $\sfrac{1}{4}$ of the atomic $U$ values once
relaxation is taken into account~\cite{hourahine2007self,PhysRevB.75.045101}. We
find a best match to the CASTPT2 results for $\tilde{U}_{3d} \approx 4.6$~eV
with this approximation, with qualitatively similar features to the FLL
results.

\section{Charge transfer in a nanodiamond donor-acceptor complex}
\label{sec:nanodiamond}

One exciting application that this implementation enables is the study of the influence of the nuclear motion on light-induced charge transfer processes in donor-acceptor complexes. These kinds of systems are the building blocks for optoelectronic materials for light to electricity conversion devices as dye sensitized solar cells or organic solar cells. 
Recent experimental and theoretical studies show that coherent vibronic coupling between electrons and nuclei is of key importance for the description of the first steps of charge separation in non-covalent \cite{Falke2014} and covalently \cite{Rozzi2013} linked organic systems, and also in vertically stacked transition-metal dichalcogenide layers \cite{Long2016,Zheng2017}.
These works reveal the increasing need to take into account the nuclear motion to describe the charge transfer processes in such materials.

In a previous work, we have studied the photodynamic process in a supramolecular arrangement composed of a hydrogenated nanodiamond (C\textsubscript{190}H\textsubscript{110}) interacting with a 3,4,9,10-perylenetetracarboxylic acid diimide (PDI) molecule as acceptor. After pulse irradiation in tune with the photoexcitation of the acceptor, the system shows an ultrafast charge transfer process reaching a stable steady-state in a few tens of femtoseconds. We proposed a purely electronic reordering of the system after charge separation as the reason for the irreversibility of the process. Up to now, we had not considered the nuclear motion for the description of this ``trap-door''-like mechanism.

\begin{figure*}[hb!]
    \centering
    \subfloat[]{%
     \includegraphics[width=0.33\textwidth]{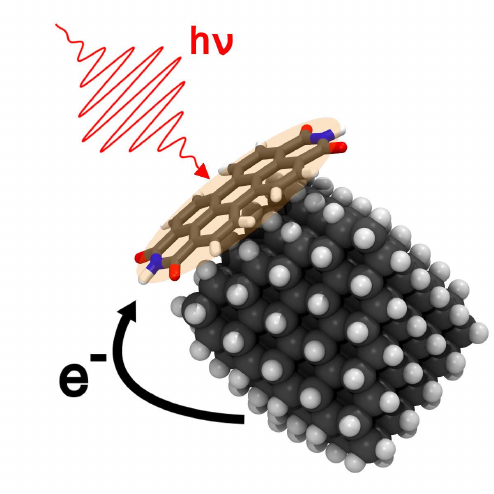}
     \label{CT-system}
    }
    \subfloat[]{%
     \includegraphics[width=0.33\textwidth]{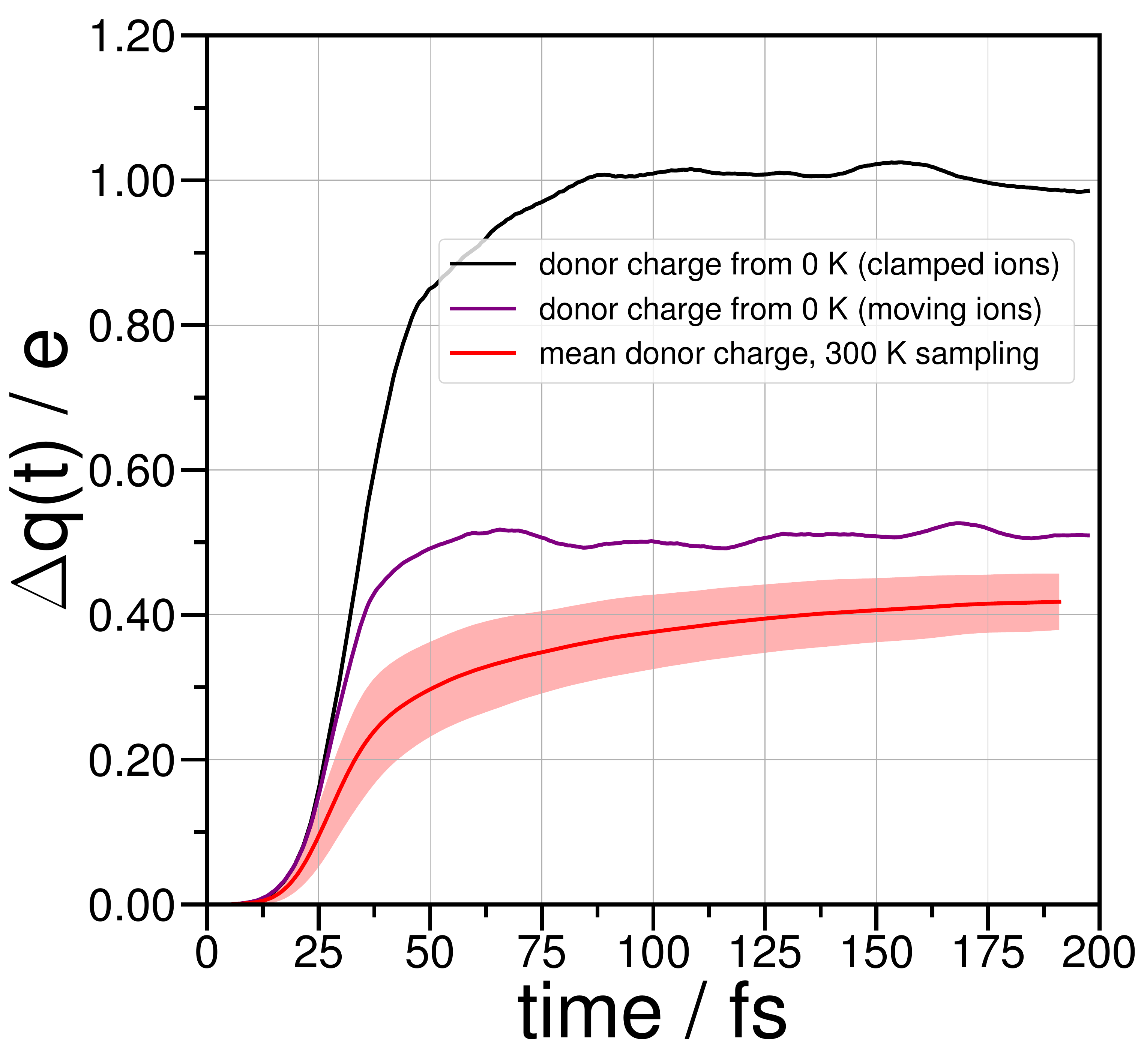}
     \label{CT-ehren}
    }
    \subfloat[]{%
     \includegraphics[width=0.33\textwidth]{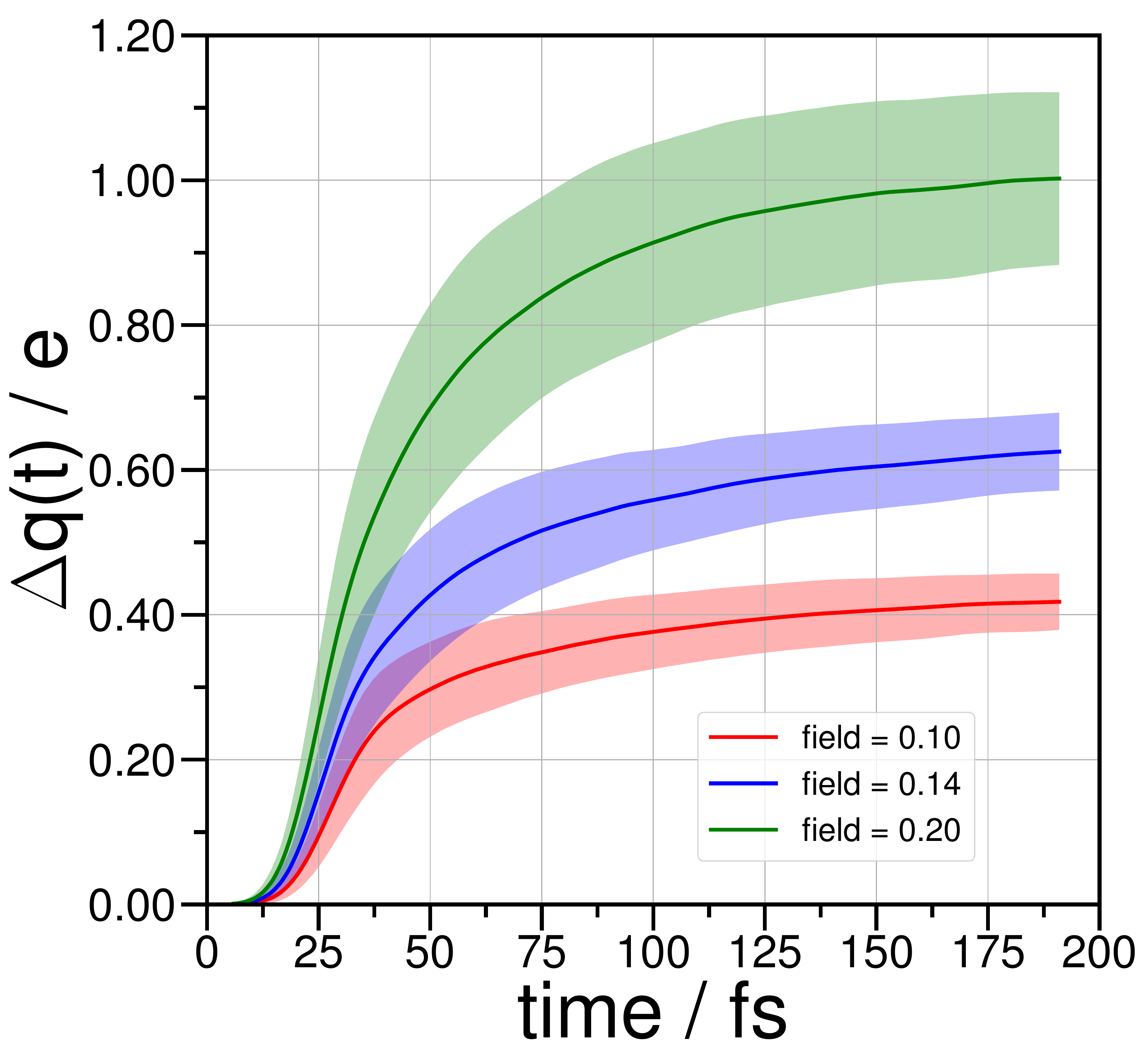}
     \label{CT-ehren-fields}
    }
    \caption{(a) C\textsubscript{190}H\textsubscript{110}+PDI donor-acceptor complex representation. The system is composed by a perylene tetracarboxilic diimide (acceptor) interacting with a hydrogen-terminated ND (donor) by London dispersion forces. The figure describes qualitatively the photodynamic process of charge transfer. (b) Donor charge as a function of time after 50 fs pulse irradiation tuned with the acceptor lower energy excitation. The red and blue curves indicate fixed (frozen) ions and moving ions, respectively.}
    \label{fig:CT_1}
\end{figure*}

Based on our experience of how systems react to sudden changes in electronic structure, we expect coherent breathing oscillations to be launched impulsively in the nanodiamond \cite{Bonafe2017a}. Here we show some preliminary simulations of the same system presented in ref.~\citenum{Medrano2018} (see figure \ref{CT-system}) including the nuclear motion using the Ehrenfest TD-DFTB method described in the section \ref{sec:theory}. 

Figure \ref{CT-ehren} shows the evolution of the Mulliken charge distribution in the ND (donor) upon the photoexcitation of the complex with a pulsed electric field perturbation in tune with the acceptor excitation energy. \revtwo{The black and purple curves represent the dynamics considering clamped and moving ions, respectively. The nuclear motion decreases the total steady-state charge transfer by half. To take into account the effect of the temperature we have also performed Ehrenfest dynamics starting from configurations sampled at 300 K from an NVT ensemble by means of Born-Oppenheimer molecular dynamics performed using the DFTB+ package~\cite{Hourahine2020} (see supporting information for the details).
Figure 3b also shows the mean charge of the donor obtained from averaging the simulations for 15 decorrelated configurations (red curve), together with its standard deviation. Higher temperatures also decrease the net charge transfer, but with a less pronounced effect.}

\revtwo{To find whether the nuclear motion limits the total charge transferred, we have also studied its dependence on the external field strength. As shown in figure 3c, charge transfer can be enhanced by using a stronger external field, reaching a total charge transfer of 1 e as in the case of fixed nuclei by increasing the field strength from 0.1 V/\AA ~to 0.2 V/\AA.}

\revtwo{In summary, under the same illumination conditions, a strong decrease of the charge transfer is observed when nuclear motion is allowed, supporting our previous hypothesis \cite{Medrano2018}. This can be interpreted as follows: nuclear motion increases the detuning, which allows the system to reach the steady-state regime sooner. This, in turn, reduces the total charge transferred, since a new relaxation channel (i.e.~the nuclear degrees of freedom) is added. However, nuclear motion and temperature do not affect the trap-door like mechanism, as the system still reaches a steady state after the pulse perturbation.} These preliminary results show that nuclear motion is not negligible even at very short time scales, having an effect in ultrafast charge transfer mechanisms.

\section{Transient absorption spectra simulations}

\subsection{Theory and computational method}
\label{sec:tas}

One of the most novel applications of the present implementation is the simulation of time-resolved pump-probe spectroscopy experiments in the UV/visible range. Within these techniques, in particular broadband time-resolved transient absorption (TA) spectroscopy is one of the most widely used and can be straightforwardly obtained using the Ehrenfest real time TD-DFTB method, up to its limitations, as will be shown below.

To measure broadband TA spectra, a system is first excited using a \textit{pump} pulse tuned to a specific wavelength. After a delay time $\tau$ has elapsed, a weaker \textit{probe} pulse is applied, and the absorption at all wavelengths withing the probe bandwidth, $A(\omega, \tau)$ is obtained. This procedure can be repeated to measure the absorption at different delay times, building a time-resolved absorption. The quantity of interest is the normalized differential transmittance, $\Delta T/T$, or the transient absorbance, $\Delta A(\omega, \tau) = A(\omega, \tau) - A_0(\omega)$, which are connected via $\Delta A = - \Delta T/T$ if $\Delta T \ll T_0$, being $T_0$ the transmittance of the unperturbed system.

Several theoretical models have been developed since the 1990s to extract quantitative information about the potential energy surfaces that drive the dynamics of the system~\cite{Pollard1992}. In non-equilibrium spectroscopy, the non-linear terms of the response function (or susceptibility) of the system are relevant to compute the total polarization of the system. In isotropic media, the even-order response functions vanish due to the inversion symmetry, thus being the third-order response function of the first non-linear term in most materials. This term can be represented as a triple commutator of the dipole moment in the interaction picture with the equilibrium DM, which gives eight terms upon expansion~\cite{Mukamel1999}. These eight terms represent different time-ordering of the interactions with the external field. For the case of TA spectra, only six of these terms survive. Following this approach and fitting parameters for model systems, usually with few (more than three) electronic levels and one or two nuclear coordinates, several simulations of TA spectra have been done~\cite{Pollard1992,Finkelstein1993,Tanimura1995,Seidner1995,Domcke1997,Bratos1997,Kumar2001,Tortschanoff2002,Gelin2003,Sugisaki2007,Polli2010}. This framework, however, is not practical when a mean field Hamiltonian (like in DFT or DFTB) is used, since the complexity of the third-order response function grows very fast with the size of the system ($\mathcal{O}(N^{16})$).

A different approach suitable for \textit{ab initio} or semi-empirical atomistic simulations is based on the real-time propagation of the full molecular or nanostructured system. The work by De Giovannini \textit{et al.} introduced the idea of using the pump-dependent effective response function $\chi_{eff}$ of the system within real-time TD-DFT for the calculation of TA spectra with pulses in the attosecond regime and without considering nuclear motion due to the short time scales~\cite{DeGiovannini2013}. \revone{Following a similar idea, Boleininger \textit{et al.} were able to compute the transient core absorption spectra of polythiophene fragments over various delay times of up to $\sim 20$ fs using Ehrenfest time-dependent Gaussian tight-binding and XUV probe pulses\cite{Boleininger2017}.}

To derive an expression that provides more insight on the structure of the $\chi_{eff}$, we start by considering a simulation protocol similar to the one used to compute the ground state spectra described in section \ref{sec:gsspec}. We aim to derive a method that allows us to use the time-dependent dipole moment, $\bmu(t)$, after probing the system with a \textit{kick} $\bE_{kick}$, subtracting all other contributions that are not frequencies excited by the probe, hereafter referred to as \textit{reference} dipole moment $\bmu_{ref}(t)$:
\begin{equation}
 \bmu(t) - \bmu_{ref}(t) = \int_{-\infty}^\infty \chi_{eff}(t,\tau) \bE_{kick}(\tau) d\tau.
 \label{eq:target}
\end{equation}
Such system has been driven out of the equilibrium by a pump pulse, represented as:
\begin{equation}
 \bE_b(t) = \Ef_0^b \sin(\omega t + \phi) f(t) \be_b,
 \label{eq:pumpf}
\end{equation}
where $\Ef_0^b$ is the peak intensity, $\omega$, $\phi$ and $f(t)$ are the carrier frequency, phase and envelope function, and $\be_b$ the polarization direction. The pump pulse total duration is $t_p$, and for simplicity we consider that it is non-zero only$0 \leq t \leq t_p$.

The expansion of the dipole moment in terms of the pump field is given by:
\begin{equation}
\begin{split}
 \bmu_b(t) =& \bmu_0^b + \int \chi^{(1)}(t,\tau)\bE_b(\tau) d\tau +
 \frac{1}{2!}\iint \chi^{(2)}(t,\tau,\tau') \bE_b(\tau)\bE_b(\tau') d\tau d\tau' +  \\
 & \frac{1}{3!}\iiint \chi^{(3)}(t,\tau,\tau',\tau'') \bE_b(\tau)\bE_b(\tau')\bE_b(\tau'') d\tau d\tau'd\tau'' + \dots
\end{split}
\label{eq:muexppump}
\end{equation}
where $\bmu_0^b$ is the static dipole moment and $\chi^{(n)}$ is the $n$-th order response function. Since the pump pulse is strong, this expansion cannot be truncated to first order, so all orders must be considered. Now, consider that at $t=t_s$ the probe pulse is applied, with the aforementioned Dirac-delta shape:
\begin{equation}
\bE_s(t)= \Ef_0^s \delta(t-t_s) \be_s,
\label{eq:probef}
\end{equation}
where $\Ef_0^s$ and $\be_s$ have the equivalent meaning as in the pump pulse. Hence, the total applied field is $\bE(t) = \bE_b(t) + \bE_s(t)$. The dipole response for the total field is then:
\begin{align}
 \bmu(t) =& \bmu_0 + \int \chi^{(1)}(t,\tau) (\bE_b(\tau) + \bE_s(\tau)) d\tau \nonumber \\ & + \iint \chi^{(2)}(t,\tau,\tau') (\bE_b(\tau) + \bE_s(\tau))(\bE_b(\tau') + \bE_s(\tau')) d\tau d\tau' \nonumber \\ & +
 \iiint \chi^{(3)}(t,\tau,\tau',\tau'') (\bE_b(\tau) + \bE_s(\tau))(\bE_b(\tau') + \bE_s(\tau'))(\bE_b(\tau'') + \bE_s(\tau'')) d\tau d\tau'd\tau'' \nonumber \\ & + \dots
\label{eq:muexptot}
\end{align}
After tedious but straightforward algebra, and neglecting terms of second-order and above in the probe pulse, eq.~\eqref{eq:muexptot} can be re-written in a series of the ground-state response functions, where the probe pulse appears in the integrand only to first order (see appendix \ref{sec:effectivechi}). This allows us to integrate over all pump-dependent variables, yielding the following formula for the dipole response:
\begin{equation}
 \bmu(t) - \bmu_b(t) = \int d\tau \chi_{eff}[\bE_b](t,\tau) \bE_s(\tau).
 \label{eq:chieff}
\end{equation}
As can be readily observed, this expression is equivalent to the desired expression \ref{eq:target}, where $\bmu_{ref} = \bmu_b$: the reference dipole moment that must be subtracted to eliminate all spurious signals is the dipole moment of a simulation with the pump only. This reference also been used in previous works~\cite{DeGiovannini2013,Fischer2015}. As can be seen in the appendix \ref{sec:effectivechi}, the pump-dependent response function $\chi_{eff}[\bE_b]$ includes its dependence to all orders on the pump pulse, keeping in mind that the linear response approximation must be valid on the probe pulse. $\chi_{eff}[\bE_b]$ is equivalent to the effective response functions used in references~\citenum{DeGiovannini2013} and~\citenum{Walkenhorst2016}.

Fourier transforming eq.~\eqref{eq:chieff} and making explicit the dependence on the probe time $t_s$:
\begin{equation}
 \bmu(\omega, t_s) - \bmu_b(\omega, t_s) = \bpol_{eff}[\bE_b](\omega,t_s) \Ef_0^s,
 \label{eq:alphaeff}
\end{equation}
where $\bpol_{eff} = \mathcal{F}(\chi_{eff})$ is the effective dynamic polarizability for each probe time $t_s$, which by assumption of the origin of the pump pulse, here is equivalent to the delay time $\tau$. Finally, we arrive at an expression like eq.~\eqref{eq:sigma} but as a function of frequency and delay time:
\begin{equation}
 \sigma(\omega,\tau) = \frac{4\pi\omega}{c}\mathrm{Im}(\alpha_{eff}[\bE_b](\omega,\tau)),
 \label{eq:sigmaeff}
\end{equation}
where $\alpha_{eff} = \frac{1}{3}\Tr[\bpol_{eff}]$. After subtracting the absorption cross section of the GS, $\sigma(\omega,\tau)$ can be compared directly with experimental TA signals.

Of course, one limitation of this approach is the absence or incorrect description of electronic decay or rise lifetimes, which is one of the most interesting features that can be extracted from experimental TA signals. However, in a recent publication we have shown for ZnTPP that important spectral features, such as the position of the ground state bleaching and excited state absorption signals, do agree with experimental data~\cite{Bonafe2018}, which opens the door to the computational assignment of TA bands from atomistic simulations. Besides, the electronic EOM could be modified to include a damping term to account for the electronic lifetimes.

A known shortcoming in these kinds of techniques is the unphysical shift of electronic resonances in the simulated TA signals, due to the lack of memory in the adiabatic approximation for the exchange-correlation (XC) functional~\cite{Elliott2012,Fuks2015}. The maximum frequency shift that can be observed in a particular excitation for each system depends only on the coupling of the monoelectronic transitions among the occupied molecular orbitals, which would be decoupled with a proper XC functional. This is also the reason for the wrong behavior in charge transfer processes upon resonant excitation in TD-DFT~\cite{Fuks2016}. Nevertheless, a detailed investigation on the peak shifting has shown that its magnitude depends on the fraction of the excited population, which makes it important for small systems~\cite{Provorse2015}. Since TD-DFTB allows to treat large molecules and nanostructures for these kind of simulations, where the excited population for a reasonable laser pump pulse is not significant with respect to the GS population, this effect is normally negligible~\cite{Bonafe2018,Hernandez2019}.

On the other hand, this method has the advantage of not being limited to the so called non-overlapping regime (when the pump and probe pulses do not overlap), but it can also be used to probe the system while the pump pulse is acting. This regime is not usually accessible by the methods reported in the literature~\cite{Seidner1995}. Of course, in the overlapping regime, the probe trajectory needs to include also the pump fields, for the response to be consistent with the pump-only trajectory.

Eqs.~\eqref{eq:alphaeff} and \eqref{eq:sigmaeff} provide a recipe for the numerical simulations. Two trajectories, one with the effect of the pump only and one with the probe acting after a delay time $\tau$, must be used to compute a single TA spectrum. This procedure must be done for all desired delay times $\tau$. While the pump pulse is added to the Hamiltonian via eq.~\eqref{eq:hsccvext}, the Dirac delta probe pulse is applied to the (excited) DM analytically~\cite{Negre2010a}. In practice, we simulate only one pump-only trajectory, and branch from it several probe trajectories, since the latter are usually shorter (50-100 fs) than the pump trajectory (between 500 fs and 1 ps in our simulations). Therefore, the probe pulse trajectories are \textit{embarrassingly parallel}. Subtracting the corresponding dipole signals, we obtain $\Delta \mu(t,\tau)$, which is damped to account for a finite electronic lifetimes using an arbitrary damping parameter $\tau_{\mathrm{damp}}$: $ \Delta \mu'(t) = \Delta \mu(t) \exp(-t / \tau_{\mathrm{damp}})$.

\begin{figure*}[hb]
  \includegraphics[width=.8\textwidth]{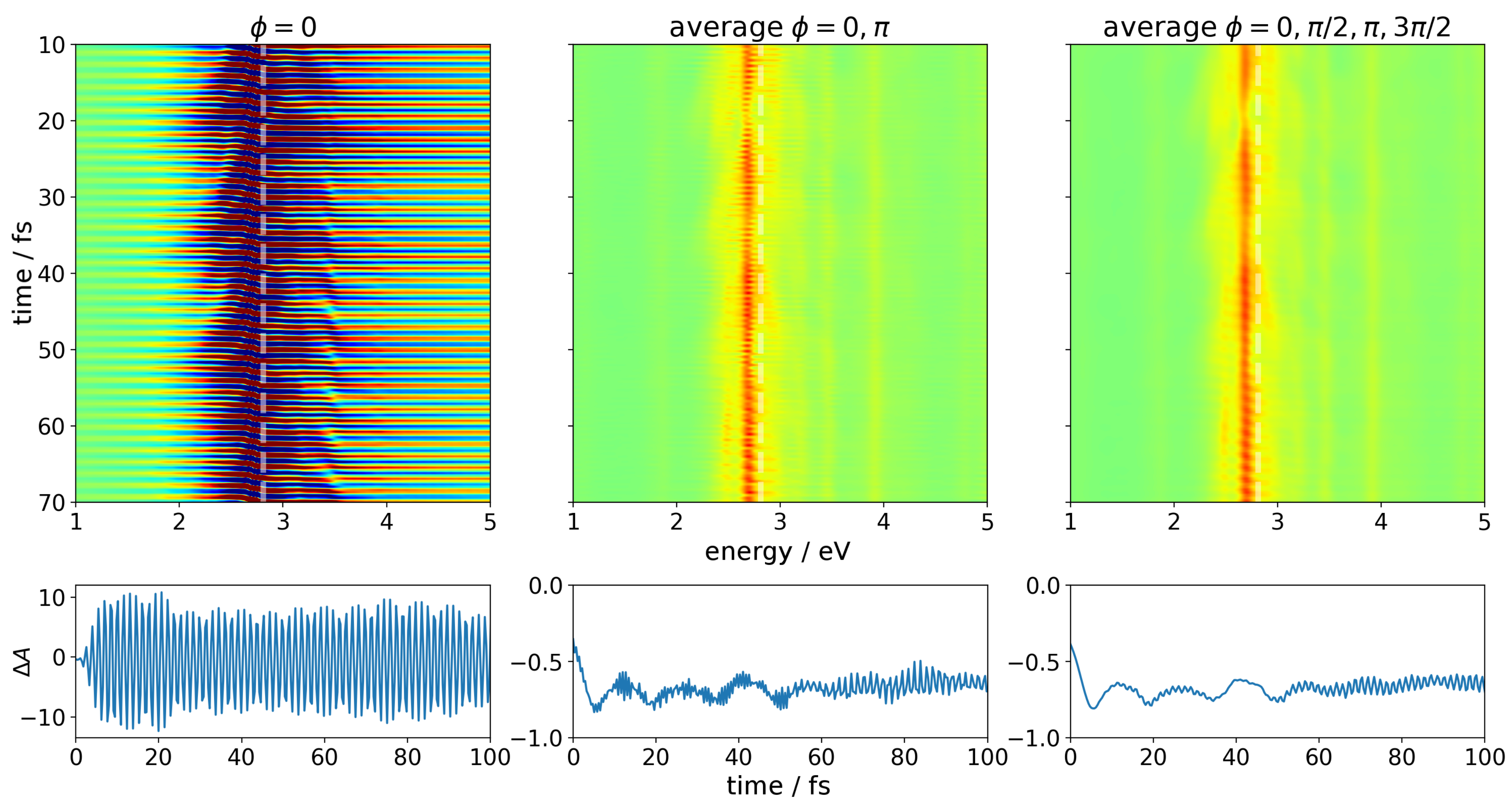}
  \caption{\label{fig:ppave} Transient absorption of ZnTPP (above) and time-traces at $\hbar \omega = 2.81$ eV  using different number of phase differences for the average: only one phase (no average) (left), two phases (middle), four phases (right).}
\end{figure*}

Last but not least, classical coherent fields leave contributions in the dipole signal of eq.~\eqref{eq:chieff} that come from the fixed relative phase between the pump and probe pulses. In experiments these contributions are not observed since the signal is averaged incoherently over several pulses~\cite{Seidner1995}, unless a phase-locked experiment is done~\cite{Cho1992}. By considering that the pump generates a superposition of eigenstates of the many-body Hamiltonian, Walkenhorst \textit{et al.} showed how the initial phases of the eigenstates affect the effective response function, making the polarizability oscillate between Lorentzian and Rayleigh peaks in time, depending on the initial phases and the energy differences between the eigenstates. Thus, if this effect is not considered, the TA signal is superimposed in an oscillating signal, in which the frequency is that of the energy difference between the states involved in the pumped transition, as it can be seen in the figure~\ref{fig:ppave} (left) for ZnTPP after excitation of the Soret peak with a 10-fs pump pulse. The upper panel shows the obtained TA signal, where all interesting features around the Soret energy are masked by this effect, while the lower panel shows the strong oscillatory behavior at the Soret energy.

\revtwo{To remove the effect of the phase difference, an average of the polarizabilities with four phase shifts between pump and probe, from 0 to 2$\pi$, must be done, as shown in ref.~\citenum{Seidner1995}. Each phase difference, that actually is assigned to $\phi$ in eq.~\eqref{eq:pumpf}, demands a different simulation of the pump and all the corresponding probes. To see the convergence of the results when increasing the number of phases in the average, in figure \ref{fig:ppave} (middle, right) the TA spectra of ZnTPP using two and four phases, respectively, are shown.} Even when the phase average multiplies the computational cost by the number of phases, each of the sets of simulations for each phase is independent of the others, and thus can be perfectly parallelized in a distributed system. Also, within TD-DFTB even for large molecular systems or small nanostructures, the computational cost is low which makes feasible such a simulation scheme.

When the phase average is not enough to get rid of all oscillatory signals, a Fourier filtering of the spectra can be done to remove all high frequencies from the signal. For example, a Gaussian filter can be applied, such that the effective polarizability is multiplied in Fourier space by a suitable Gaussian function, eliminating all the amplitude at large frequencies, and the filtered polarizability is then reconstructed by an inverse Fourier transform.

\subsection{Ultra-fast dynamics of ZnTPP on- and off-resonance}
\label{sec:znttp}

In a previous work we have presented the TA spectra of zinc(II) tetraphenylporphyrin (ZnTPP) after Soret-band excitation \cite{Bonafe2018}. In this section we present results applying the methodology to calculate TA spectroscopy, shown in section \ref{sec:tas}, for on- and off-resonance excitation of ZnTPP, and we show how it allows to distinguish the modes that are mostly coupled to the electronic transition.

\begin{figure*}
  \includegraphics[width=.9\textwidth]{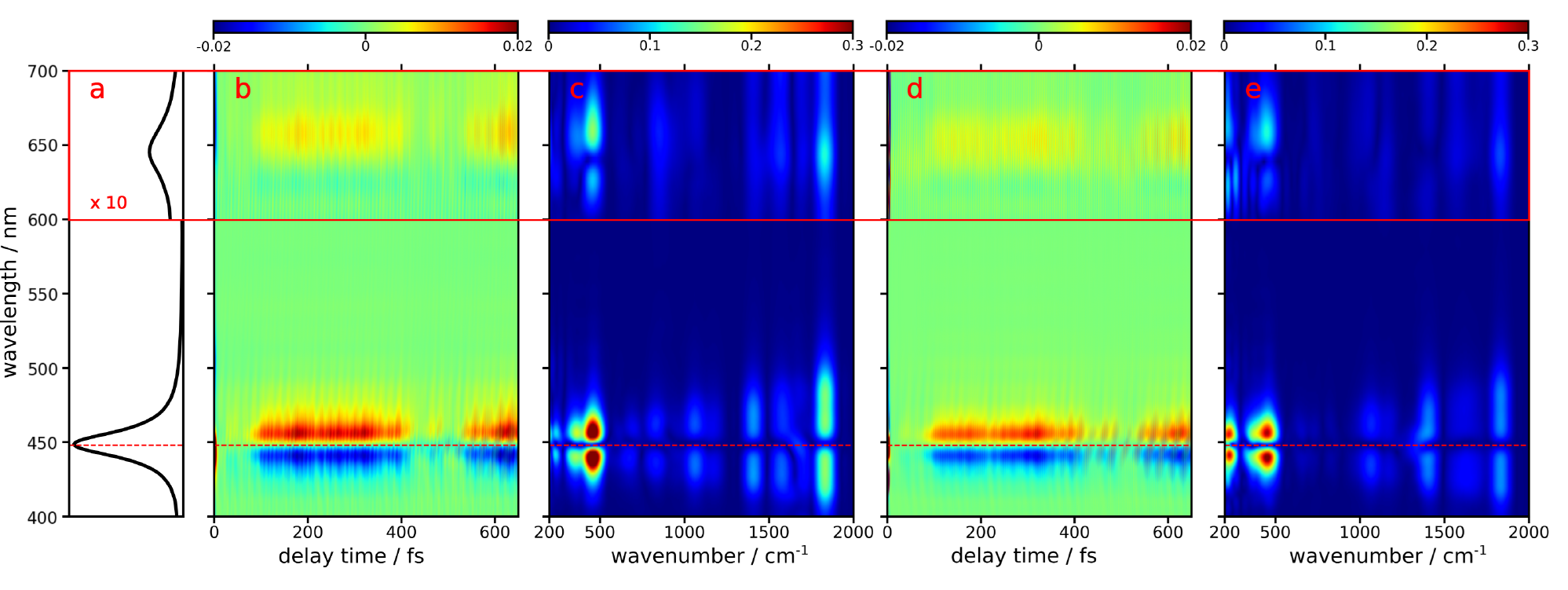}
  \caption{\label{fig:TAS_ZnTPP} ZnTPP's ground state spectrum (a); Transient absorption spectra calculated by subtracting the absorption spectrum at each time with the spectrum at the time of the pulse for the pump pulse in resonance with the Soret band at 448 nm (b) and detuned to the red at 504 nm (d). Fourier transform of the TA spectra along the delay time axis for the TA calculated in resonance with the Soret Band (c) and detuning to the red (e). The dashed line indicates the electronic excitation (pump) wavelength.}
\end{figure*}

First, the GS absorption spectrum is calculated by real-time propagation after a Dirac delta perturbation, as it is shown by Eqs. \ref{eq:alpha} and \ref{eq:sigma}. The computed absorption spectrum is presented in figure \ref{fig:TAS_ZnTPP}a. The two lowest-energy active bands, the Q-band at 1.93 eV (642 nm) and the Soret or B band at 2.77 eV (448 nm), are in relatively good agreement with the experimental values of 594 and 406 nm, respectively \cite{Edwards1971b}.

Then, ZnTPP is electronically excited by a monochromatic $\sin^2$ pulse tuned to the Soret band. The pulse has a duration of $t_p = 10$ fs and peak field intensity of 0.02 V \AA$^{-1}$ (pump). The excited system evolves for 725 fs. As it is explained in section \ref{sec:tas}, 1000 snapshots of the DM and geometry of the system are stored to obtain the TA spectra with a time resolution of 0.72 fs. Three trajectories after Dirac-delta \textit{kicks} are run for each one of the four pump trajectories (with different phase differences) giving a total amount of 12000 trajectories. It is worth mentioning that in this case the TA spectra are not calculated subtracting the GS spectrum, as it is usually done, since the strong bleach at the Soret band obscures the subtle changes that occur after the pump has acted. Therefore, the absorption change is calculated with respect to the spectrum at the end of the pump as shown by eq. \ref{eq:deltaa}:
\begin{equation}
\Delta A (t) = A(t) - A(t=t_p).
\label{eq:deltaa}
\end{equation}

Fig. \ref{fig:TAS_ZnTPP}b and \ref{fig:TAS_ZnTPP}d show the TA spectra computed with the pump pulse tuned to the Soret band of ZnTPP at 2.77 eV (on-resonance spectra) and with the pump pulse centered at 2.46 eV, i.e.~detuning the pump laser by 0.31 eV to the red of the Soret band (off-resonance spectra), respectively. As can be observed in both figures the signal presents oscillatory features with changes in the sidebands of the Soret band but not in the peak center and the intensity of such oscillations is higher when the pump pulse is tuned in resonance with the Soret band. In order to get further information of these oscillatory features, the TA spectra are Fourier transformed along the delay time axis, obtaining the spectral densities shown in figures \ref{fig:TAS_ZnTPP}c and \ref{fig:TAS_ZnTPP}e for the on-resonance and off-resonance spectra, respectively. Such spectral densities relate the absorption energy of the transient species and the energy of the different vibrational motions coupled to the ground-state absorption band that is pumped. Hence, all observed peaks in the region between 200 and 2000 cm$^{-1}$ in figures \ref{fig:TAS_ZnTPP}b and \ref{fig:TAS_ZnTPP}d, constitute the fingerprint of vibrational modes that are impulsively activated by the pump pulse. The spectral density is finally integrated within the pumped absorption peak region, yielding a spectrum which contains the Raman activity of the system, providing similar information to that available in the frequency domain.

 This computational methodology simulates a state-of-the-art spectroscopic technique known as vibrational coherence spectroscopy (VCS) and the type of spectrum thus obtained is commonly called impulsive vibrational (IV) spectrum \cite{Collini2010,Scholes2011,Huelga2013,Fuller2014,Romero2014,Scholes2017,Gueye2018}. VCS's main advantage over frequency domain Raman spectroscopy is that the full vibrationally coherent evolution of the system is measured. This allows to track molecular structural dynamics accompanying ultrafast photoinduced processes in molecular systems. However, VCS has the drawback that vibrational coherences in both ground and excited states might be present and disentangling them can be challenging. 
 
 In previous works,\cite{Bonafe2018,Hernandez2019} we have proposed interpreting the IV spectrum along with the potential energy distributed (PED) on ground-state normal modes after the pulse. The PED is obtained from the pump-only trajectory, \revtwo{starting from the equilibrium geometry with the atoms at rest,} and projecting the resulting time-resolved nuclear displacements onto normal modes coordinates as follows \cite{Horiuchi1991}:
\begin{equation}
Q_i (t) = \sum_A m_A \Delta\mathbf{r}_A(t) \cdot \mathbf{v}_{Ai},
\label{eq:modesprojection}
\end{equation}
where $\mathbf{v}_{Ai}$ are the eigenvector matrix elements, $\Delta \mathbf{r}_A(t)$ are the nuclear displacements, $m_A$ is the atomic mass and $Q_i$ are the coordinates in the normal mode basis set. The potential energy in the $i$-th mode can be calculated as shown in eq. \ref{eq:modespotenergy}, where $c$ is the speed of light and $\bar{\nu}_i$ the mode's wavenumber:
\begin{equation}
V_i (t) = (2 \pi c \bar{\nu}_i Q_i(t))^2.
\label{eq:modespotenergy}
\end{equation}

A major problem when analyzing the IV spectrum of a molecule with a large number of normal modes, such as ZnTPP (225 normal modes), is discriminating the vibrational coherences directly coupled to the electronic transition from those which are not. The solution we propose to overcome this issue is to compute two IV spectra with the exact same conditions but different pump frequency: one in resonance with the selected electronic transition and the other off-resonance (detuned to the red in this case). The differences in both IV spectra can thus be attributed to the vibrational coherences directly coupled to the electron transition and therefore, the other  vibrational coherences can be filtered out.

\begin{figure}
  \includegraphics[width=.55\textwidth]{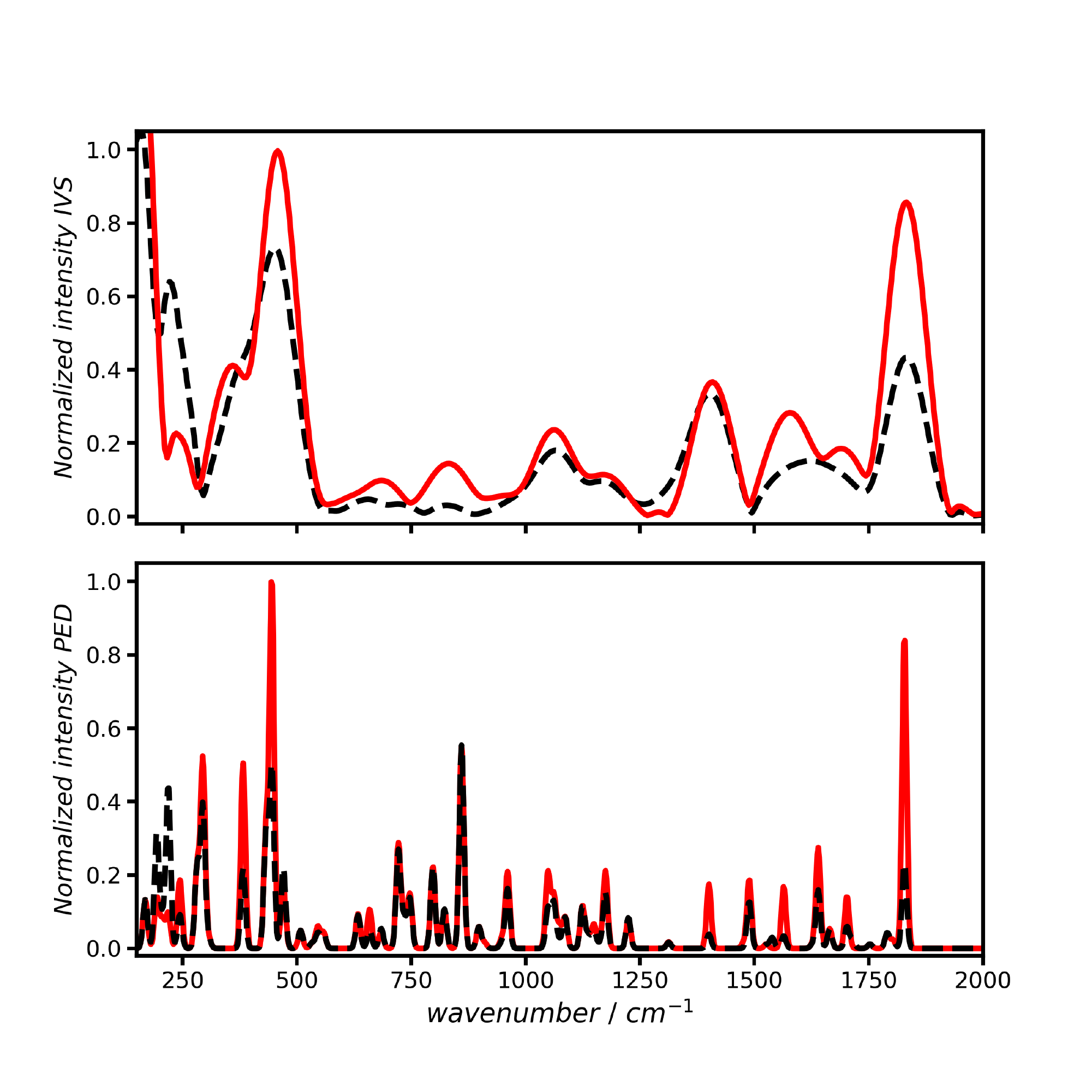}
  \caption{\label{fig:IVS_ZnTPP} Impulsive vibrational spectra (top) and potential energy distributions per normal mode (bottom) of ZnTPP. The red trace is calculated with the pump laser in resonance with the Soret band, whilst the black trace detuning the pump laser in 0.31 eV to the red.}
\end{figure}

In figure \ref{fig:IVS_ZnTPP} (top pannel) we show the IV spectra computed on and off-resonance after the integration of the spectral densities displayed in figures \ref{fig:TAS_ZnTPP}c and \ref{fig:TAS_ZnTPP}e, respectively, in comparison to the corresponding PEDs (bottom pannel). 
We have studied the good correlation between the IV spectrum and the PED in previous works \cite{Bonafe2018,Hernandez2019}. What is different in this case, as shown in figure \ref{fig:IVS_ZnTPP}, is the drop in intensity in both the IV spectrum and the PED for the peaks centered in 448 and 1830 cm$^{-1}$. We have observed that these modes are the most coupled with the Soret transition \cite{Bonafe2018}, in a partial agreement with experimental evidence \cite{Abraham2016c, Yoon2003d}. These modes strongly modify the energy gap of the illuminated absorption band \cite{Liebel2014} and, as we have shown, can be discriminated by direct comparison of the response at different pump energies by using the Ehrenfest TD-DFTB method. \revtwo{In summary, the calculation of the less computationally expensive PED does not replace the information provided by the simulated IV spectrum, but rather helps to interpret, identify and understand what normal modes are modulating the TA signal and correlate it with the symmetry of the transition density~\cite{Bonafe2018}. The comparison of on-resonance and off-resonance simulations, enables us to further identify which modes are more strongly driven by the external laser.}

\revtwo{Finally, some details about the computational resources and runtime of the calculations for this section should be discussed. Running each simulation on 4 Intel Xeon E5-2670 cores, one \textit{pump} trajectory (600000 steps) takes 6.5 hours, while each \textit{probe} (3 runs of 40000 steps each) takes 25 minutes. Since the latter are perfectly parallel, they can be run on a large number of nodes simultaneously. Using only eight machines with 16 cores each, the calculation of all 24000 probe runs required to compute the TA spectra took 4 days in total.}

\section{Conclusions}
\label{sec:conclusions}

In this work, we have explained the details of a novel implementation of electron-nuclear Ehrenfest TD-DFTB dynamics in the DFTB+ package and some important applications. We have developed a versatile, free and open-source code to study photoinduced proceses considering nuclear motion in molecular, nanoscopic and periodic systems. The use of a semiempirical method as the base formalism increases considerably the efficiency of this implementation in comparison with ab-initio methods for real-time dynamics. The ability of this tool to tackle problems previously thought to be out of the reach of atomistic quantum dynamics methods, when modest computational resources are available, has been demonstrated. First of all, by having at our disposal the underlying infrastructure developed in DFTB+ for ground state calculations, the range of possible systems that can be studied is amplified. This includes systems with correlated localized electrons through the ``DFTB+U'' correction, and systems that are periodic in one, two or three dimensions. Simulations of the optical properties for some of these systems have been discussed and in some cases, they even show improvements with respect to TD-DFT. Moreover, the coupling of nuclear motion with electronic processes such as the ultrafast electron transfer in supramolecular arrangements is discussed, in this case with a structure comprising 340 atoms. A change in the charge transfer mechanism is evident by the decrease in the charge that is transferred after excitation of the acceptor. Since this is an indirect charge transfer process, the relaxation channels that the nuclear motion opens has a dramatic effect in the efficiency of the process. This can be captured at the Ehrenfest level which makes our implementation a suitable tool for future studies in the field of novel photovoltaic systems such as organic and hybrid solar cells.

Among the most important applications of this implementation is the simulation of time-resolved and frequency-resolved excited-state spectroscopies within the limitations of Ehrenfest dynamics (poorly described electronic relaxation and no electron-phonon inelastic scattering). Here, the fundamentals of transient absorption spectra simulations and its application to compute impulsive vibrational spectra are shown. While the electronic dynamics after excitation lack an accurate description of the exponential damping, the method surprisingly describes well the transient absorption signals as has been discussed in ref.~\citenum{Bonafe2018} and the vibrational signature on the spectral signal agrees with experimental measurements. This is justified in short time scales where the electron-phonon dissipation has not developed yet, and hence the Ehrenfest method is able to capture the electron-nuclear interactions in the quasi-adiabatic excited-state regime after the pulse irradiation. This tool will most certainly be of help to the scientific community for elucidating the atomistic mechanisms in light-induced processes of relevant systems.

Regarding future developments in the tool, there are several challenges that still need to be addressed. Time-dependent perturbations along a direction in which the system is periodic would be an important feature for our tool. Higher-order corrections to the Ehrenfest methods, such as the coupled electron-ion dynamics method, can be added to improve the description of the electron-nuclear correlation, thus increasing the power of the tool to describe photochemical processes and other phenomena such as Joule heating. Moreover, the addition of open boundaries to study the dynamics of systems coupled to reservoirs will make this a new powerful tool in the field of excited-state molecular electronics. In parallel to these additions, the performance of the tool will be enhanced and the real-time dynamics of systems of more than 10000 atoms will be within reach using distributed computing systems. With these further additions and others, we expect that this tool will open new doors for exciting discoveries in several research fields related to light matter-interactions at the molecular and nanoscopic level.

\begin{acknowledgement}
 F.P.B., C.R.M. and F.J.H. thank CONICET for the doctoral and postdoctoral fellowships. The authors thank DFG-RTG2247 for the funding and the BCCMS for the computational resources provided. C.G.S. is grateful for support provided by Consejo Nacional de Investigaciones Científicas y Técnicas (CONICET, Argentina) through grant PIP 11220170100892CO, and Agencia Nacional de Promoción Científica y Tecnológica (ANPCYT, Argentina) through grant PICT-2017-1506. The authors thank Andrew Horsfield for the insightful discussions.
\end{acknowledgement}

\begin{suppinfo}
Details regarding the thermal sampling of configurations for section \ref{sec:nanodiamond} are included in the Supporting Information.
\end{suppinfo}

\appendix
\section{Derivation of the density matrix EOM}
\label{sec:derivationeom}
To simplify the equations, the NAC can be thought of as an effective modification to the Hamiltonian. Let us define:
\begin{equation}
 \tilde{H}_{\mu \nu} = H_{\mu \nu} - \im D_{\mu \nu}.
\end{equation}

Now, to derive the EOM for the density matrix and starting from the density matrix operator $\hat{\rho}$:
\begin{equation}
\rho_{\nu \mu} = \sum_{i} C_{i\nu} f_i C^*_{i \mu},
\end{equation}
the derivative with respect of time leads to:
\begin{align}
 \dot{\rho}_{\mu \nu} &= \sum_i f_i (\dot{C}^*_{i\nu} C_{i\mu} + C^*_{i\nu} \dot{C}_{i\mu}) \nonumber \\
 \dot{\rho}_{\mu \nu} &= \sum_i f_i \left\{
 \im \left(\sum_{\kappa \lambda} \invs_{\kappa \nu} \tilde{H}^*_{\kappa \lambda} C^*_{i \lambda} \right) C_{i \mu} +
 \left(-\im \sum_{\eta \xi} \invs_{\eta \mu} \tilde{H}_{\eta \xi}C_{i \xi} \right) C^*_{i \nu}
 \right\} \nonumber \\
 \dot{\rho}_{\mu \nu} &= \im \sum_i f_i \left\{
 \sum_{\kappa \lambda} \invs_{\kappa \nu} \tilde{H}^\dagger_{\lambda \kappa} C^*_{i \lambda} C_{i \mu} -
 \sum_{\eta \xi} \invs_{\eta \mu} \tilde{H}_{\eta \xi} C_{i \xi} C^*_{i \nu}
 \right\} \nonumber \\
 \dot{\rho}_{\mu \nu} &= - \im \left\{ \sum_{\eta \xi} \invs_{\mu \eta} \tilde{H}_{\eta \xi} \rho_{\xi \nu} - \sum_{\kappa \lambda} \rho_{\mu \lambda} \tilde{H}^\dagger_{\lambda \kappa}  \invs_{\kappa \nu} \right\} \nonumber \\
 \dot{\rho}_{\mu \nu} &= -\im (\invs \tilde{H} \rho - \rho \tilde{H}^\dagger \invs)_{\mu \nu}
\end{align}

But by definition $\tilde{H}$ is not necessarily hermitian: $\tilde{H}^\dagger = H +\im D^\dagger$. By substituting with this expression we arrive to the EOM of eq.~\eqref{eq:eomrho}.

\section{Effective response function}
\label{sec:effectivechi}

Starting from eq.~\eqref{eq:muexptot}, expanding the terms between parenthesis and neglecting terms of order 2 and above:
\begin{equation}
\begin{split}
 \bmu(t) =& \bmu_0 + \int d\tau  \chi^{(1)}(t,\tau) \bE_b(\tau) + \int d\tau \chi^{(1)}(t,\tau)\bE_s(\tau) +\\ &
 \frac{1}{2!}\iint d\tau d\tau'  \chi^{(2)}(t,\tau,\tau') \bE_b(\tau)\bE_b(\tau') + \\
 & \frac{1}{2!}\iint d\tau d\tau' \chi^{(2)}(t,\tau,\tau') \bE_b(\tau)\bE_s(\tau') + \iint d\tau d\tau' \chi^{(2)}(t,\tau,\tau') \bE_b(\tau')\bE_s(\tau) + \\
 & \frac{1}{3!}\iiint d\tau d\tau'd\tau'' \chi^{(3)}(t,\tau,\tau',\tau'') \bE_b(\tau)\bE_b(\tau')\bE_b(\tau'') + \\
 & \frac{1}{3!}\iiint d\tau d\tau'd\tau'' \chi^{(3)}(t,\tau,\tau',\tau'') \bE_b(\tau) \bE_b(\tau') \bE_s(\tau'') +\\
 & \frac{1}{3!}\iiint d\tau d\tau'd\tau'' \chi^{(3)}(t,\tau,\tau',\tau'') \bE_b(\tau') \bE_b(\tau'') \bE_s(\tau') +\\
 & \frac{1}{3!}\iiint d\tau d\tau'd\tau'' \chi^{(3)}(t,\tau,\tau',\tau'') \bE_b(\tau') \bE_b(\tau'') \bE_s(\tau) + \cdots
\end{split}
\label{eq:muexpdist}
\end{equation}
By inspecting eqs.~\eqref{eq:muexpdist} and \eqref{eq:muexppump}, we see that the terms that depend only on $\bE_b$ are equal. Then, all these terms can be grouped in the pump-generated dipole, $\bmu_b(t)$, and hence eq.~\eqref{eq:muexpdist} can be rewritten as:
\begin{equation}
\begin{split}
 \bmu(t) =& \bmu_b(t) + \int d\tau \chi^{(1)}(t,\tau)\bE_s(\tau) +  \frac{1}{2!}\iint d\tau d\tau' \chi^{(2)}(t,\tau,\tau') \bE_b(\tau)\bE_s(\tau') + \\
 & \frac{1}{2!} \iint d\tau d\tau' \chi^{(2)}(t,\tau,\tau') \bE_b(\tau')\bE_s(\tau) + \\
 & \frac{1}{3!}\iiint d\tau d\tau'd\tau'' \chi^{(3)}(t,\tau,\tau',\tau'') \bE_b(\tau) \bE_b(\tau') \bE_s(\tau'') +\\
 & \frac{1}{3!}\iiint d\tau d\tau'd\tau'' \chi^{(3)}(t,\tau,\tau',\tau'') \bE_b(\tau') \bE_b(\tau'') \bE_s(\tau') +\\
 & \frac{1}{3!}\iiint d\tau d\tau'd\tau'' \chi^{(3)}(t,\tau,\tau',\tau'') \bE_b(\tau') \bE_b(\tau'') \bE_s(\tau)
 + \mathcal{O}((\Ef_0^s)^2) + \dots
\end{split}
\label{eq:muexpdistshort}
\end{equation}
We recognize now that all terms of order $n>1$ can be grouped as it is shown here for the second-order term:
\begin{align*}
 \bmu^{(2)}(t) =& \frac{1}{2!} \iint_{-\infty}^\infty d\tau d\tau' \chi^{(2)}(t,\tau,\tau') \bE_b(\tau)\bE_s(\tau') + \frac{1}{2!} \iint_{-\infty}^\infty d\tau d\tau' \chi^{(2)}(t,\tau,\tau') \bE_b(\tau')\bE_s(\tau) \\
 =&\frac{1}{2!} \int_{-\infty}^\infty d\tau' \left( \int_{-\infty}^\infty d\tau \chi^{(2)}(t,\tau,\tau') \bE_b(\tau)) \right) \bE_s(\tau')\\ & + \frac{1}{2!} \int_{-\infty}^\infty d\tau \left( d\tau' \chi^{(2)}(t,\tau,\tau') \bE_b(\tau') \right) \bE_s(\tau) \\
 =&\frac{1}{2!} \int_{-\infty}^\infty d\tau' \left( \int_{-\infty}^\infty d\tau \chi^{(2)}(t,\tau,\tau') \bE_b(\tau)) \right) \bE_s(\tau')\\ &+
 \frac{1}{2!} \int_{-\infty}^\infty d\tau \left( \int_{-\infty}^\infty  d\tau' \chi^{(2)}(t,\tau,\tau') \bE_b(\tau') \right) \bE_s(\tau)
\end{align*}

As $\chi^{(2)}(t,\tau,\tau') = \chi^{(2)}(t,\tau',\tau)$~\cite{Parr1989}, the factors between parenthesis are equal (since the integrals run from $-\infty$ to $\infty$), which makes both integrals equal. Therefore:
\begin{equation}
\bmu^{(2)}(t) = 2 \frac{1}{2!} \int_{-\infty}^\infty d\tau \chi_{eff}^{(2)}[\bE_b](t,\tau)  \bE_s(\tau).
\end{equation}
Similarly,
\begin{equation}
 \bmu^{(3)}(t) = 3\frac{1}{3!} \int_{-\infty}^\infty d\tau \chi_{eff}^{(3)}[\bE_b](t,\tau)  \bE_s(\tau),
\end{equation}
and in general
\begin{equation}
 \bmu^{(n)}(t) = n\frac{1}{n!} \int_{-\infty}^\infty d\tau \chi_{eff}^{(n)}[\bE_b](t,\tau)  \bE_s(\tau).
\end{equation}

Now we can rewrite eq.~\eqref{eq:muexpdistshort} as:
\begin{equation}
\begin{split}
 \bmu(t) =& \bmu_b(t) + \int d\tau \chi^{(1)}(t,\tau)\bE_s(\tau) + 2 \frac{1}{2!} \int_{-\infty}^\infty d\tau \chi_{eff}^{(2)}[\bE_b](t,\tau)  \bE_s(\tau) + \cdots \\
%  3\frac{1}{3!} \int_{-\infty}^\infty d\tau \chi_{eff}^{(3)}[\bE_b](t,\tau)  \bE_s(\tau) + \mathcal{O}((\Ef_0^s)^2) + \dots \\
 \bmu(t) =& \bmu_b(t) + \sum_{n=1}^\infty \frac{1}{n!} n \int d\tau \chi_{eff}^{(m)}[\bE_b](t,\tau)\bE_s(\tau) + \mathcal{O}((\Ef_0^s)^2) \\
 &= \bmu_b(t) + \int d\tau \left(\chi^{(1)}(t,\tau) + \sum_{n=2}^\infty \frac{1}{(n-1)!} \chi_{eff}^{(m)}[\bE_b](t,\tau)\right) \bE_s(\tau) + \mathcal{O}((\Ef_0^s)^2) \\
 &= \bmu_b(t) + \int d\tau \chi_{eff}[\bE_b](t,\tau) \bE_s(\tau) + \mathcal{O}((\Ef_0^s)^2)
\end{split}
\label{eq:muexpeff}
\end{equation}
Rewriting this like eq.~\eqref{eq:chieff}, one arrives at the desired expression for $\chi_{eff}$ in eq.~\eqref{eq:chieffdef}:
\begin{equation}
\begin{split}
\chi_{eff}[\bE_b](t,\tau) &= \chi^{(1)}(t,\tau) + \sum_{n=2}^\infty \frac{1}{(n-1)!} \chi_{eff}^{(m)}[\bE_b](t,\tau) \\
&= \chi^{(1)}(t,\tau) + \int d\tau' \chi^{(2)}(t,\tau,\tau') \bE_b(\tau')\\ & + \iint d\tau'd \tau'' \chi^{(3)}(t,\tau,\tau',\tau'') \bE_b(\tau') \bE_b(\tau'') + \dots
\end{split}
\label{eq:chieffdef}
\end{equation}

\bibliography{biblio2}

\end{document}

\section{Additional data for section 4}

\subsection{Molecular dynamics}

\begin{figure}[htb!]
    \centering
    \includegraphics[width=0.45\textwidth]{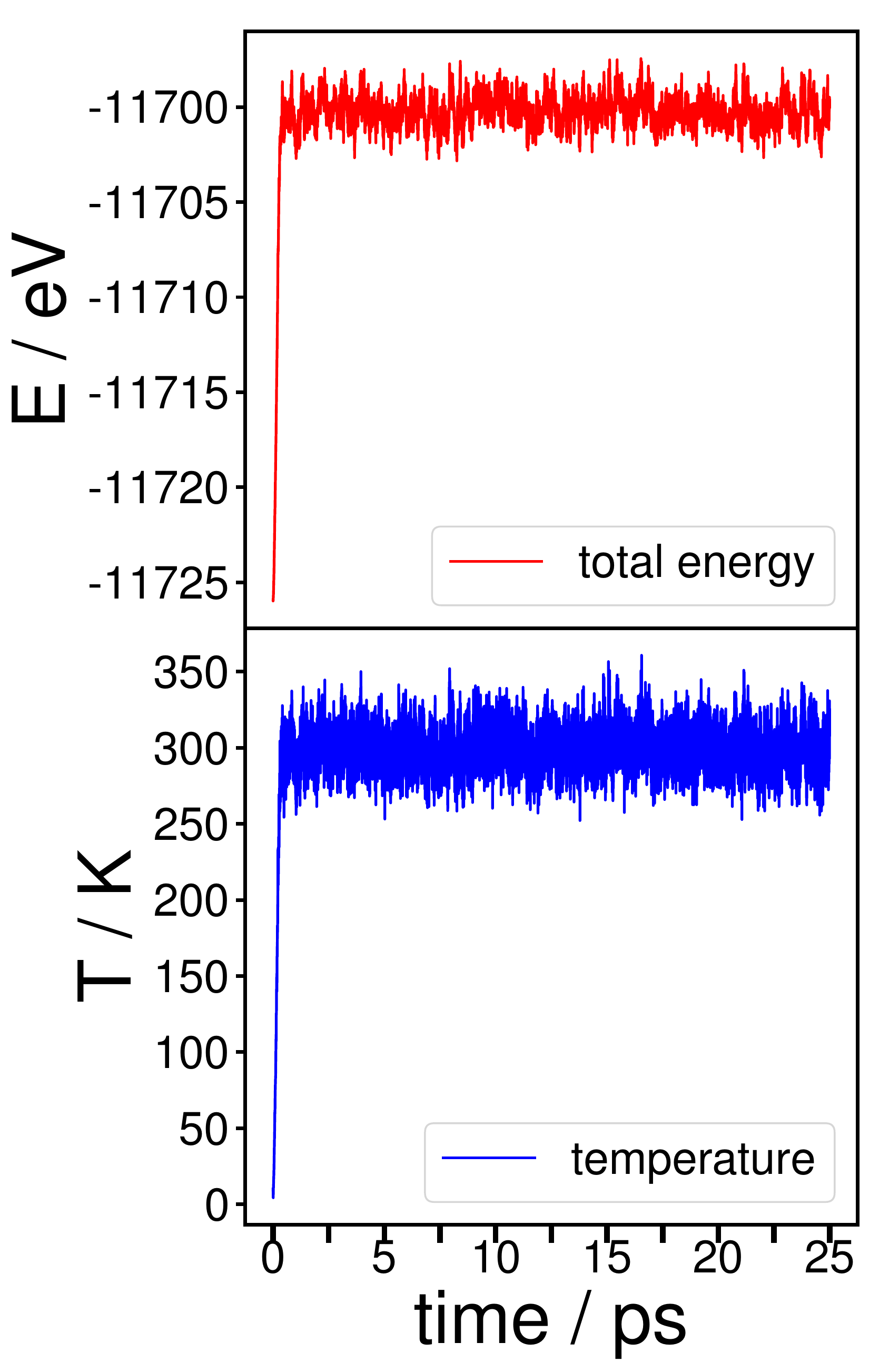}
    \caption{(up) Total energy of the system as a function of time.(bottom) Temperature of the system as a function of time.}
    \label{fig:NVT}
\end{figure}

In order to study the influence of the temperature in the charge transfer process we sample the complex configuration space using a Born-Oppenheimer molecular dynamics scheme. We have simulated 25 ps with a time step of 0.25 fs at 300 K (linearly increasing the temperature in the first 1000 steps) for the system described in the present work (C\textsubscript{190}H\textsubscript{110}+PDI), within a NVT ensemble using an Andersen Thermostat. The figure \ref{fig:NVT} shows the total energy and the temperature throughout the simulated time.

\subsection{Velocity autocorrelation function}

\begin{figure}[htb!]
    \centering
    \includegraphics[width=0.45\textwidth]{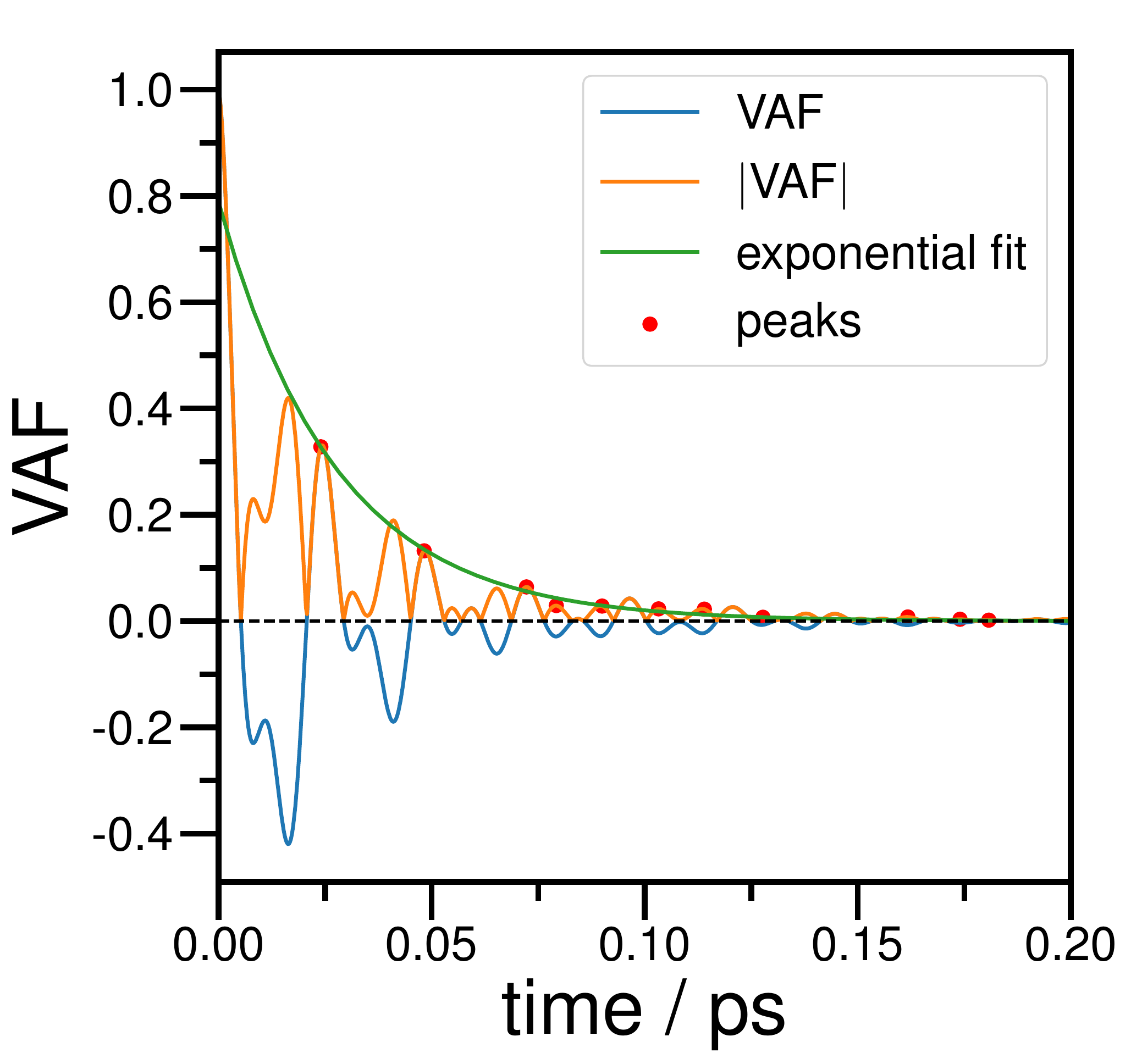}
    \caption{Velocity autocorrelation function as a function of time (blue), absolute value of the function (orange), peaks of the oscillations (red) and exponential decay fit (green). }
    \label{fig:VAF}
\end{figure}

The velocity autocorrelation function (VAF) was calculated by averaging over the entire 25 ps MD trajectory with a correlation time window of 5 fs, following previous strategies and methods\cite{Aradi2015}. In the figure \ref{fig:VAF} the correlation function as a function of time and the exponential fitting can be seen, showing the decay of the autocorrelation.

\subsection{Relevant details of the selection of frames and Ehrenfest dynamics}

To study the influence of the temperature in the charge transfer process we took 15 configurations after 6.25 ps, in intervals of 1.25 ps between them to ensure decorrelation. Note that this separation time is roughly 10 times bigger than the decorralation time calculated (See figure \ref{fig:VAF}). Then, we ran Ehrenfest dynamics for each configuration and initial velocities sampled from the molecular dynamics. The pulse perturbation was in tune with the acceptor excitation energy calculated for the optimized geometry in all cases.

\bibliography{biblio}